\documentclass[journal]{IEEEtran}
\IEEEoverridecommandlockouts

\ifCLASSINFOpdf
  \usepackage[pdftex]{graphicx}
\else  
   \usepackage[dvips]{graphicx}
\fi

\usepackage{amsmath,amsfonts}
\usepackage{amsthm}
\usepackage{algorithm}
\usepackage{array}
\usepackage{textcomp}
\usepackage{stfloats}
\usepackage{url}
\usepackage{verbatim}
\usepackage{graphicx}
\usepackage{cite}
\usepackage{cuted}
\usepackage{setspace}
\usepackage{amssymb}
\usepackage{mathtools}
\usepackage{amsthm}
\usepackage{algpseudocode}
\usepackage{makecell}
\usepackage{caption}
\usepackage{float}
\usepackage{adjustbox}
\usepackage[table,x11names]{xcolor}
\usepackage{colortbl}
\usepackage{fancyhdr}
\usepackage{optidef}
\usepackage{siunitx}
\usepackage{color}
\usepackage{tikz}
\usepackage{pgfplots}
\usepackage{subcaption}
\usepackage{tabularx}

\newtheorem{lemma}{Lemma}

\newtheorem{prop}{Proposition}
\newtheorem{remark}{Remark}

\newcommand\at[2]{\left.#1\right|_{#2}}

\pgfplotsset{compat=1.18}

\def\BibTeX{{\rm B\kern-.05em{\sc i\kern-.025em b}\kern-.08em
    T\kern-.1667em\lower.7ex\hbox{E}\kern-.125emX}}

\begin{document}

\title{Energy-Efficient Resource Allocation for PA Distortion-Aware M-MIMO OFDM System 

\thanks{The work of S. Marwaha was funded in part by BMDV 5G COMPASS project within InnoNT program under Grant 19OI22017A. Research of P. Kryszkiewicz was funded by the Polish National Science Center, project no. 2021/41/B/ST7/00136. For the purpose of Open Access, the author has applied a CC-BY public copyright license to any Author Accepted Manuscript (AAM) version arising from this submission. The work of E. Jorswieck was supported partly by the Federal Ministry of Education and Research (BMBF), Germany, through the Program of Souverän, Digital, and Vernetzt Joint Project 6G-RIC under Grant 16KISK031.
}
}

\author{Siddarth~Marwaha,~\IEEEmembership{Student Member,~IEEE,}
Pawel~Kryszkiewicz,~\IEEEmembership{Senior Member,~IEEE,}
and~Eduard~Jorswieck,~\IEEEmembership{Fellow,~IEEE}
    }

\maketitle

\begin{abstract}
While the energy efficiency (EE) of massive multiple-input multiple-output (MIMO) systems is increasingly important, it can be improved by adapting transmission parameters to the characteristics of practical radio front-ends, particularly power amplifiers (PAs). However, most existing resource-allocation strategies do not capture the nonlinear distortion–rate–power consumption interactions introduced by practical PAs, leading to inaccurate EE predictions and suboptimal designs.
To address this limitation, we study a downlink multi-user (MU) MIMO system employing orthogonal frequency-division multiplexing (OFDM) and zero-forcing (ZF) precoding under independent Rayleigh fading, where PAs are modeled using a soft-limiter with both ideal and practical power-consumption architectures. We formulate an EE maximization problem that jointly optimizes per-user transmit power and the number of active antennas.
The nonlinear PA distortion naturally restricts the feasible operating region, eliminating the need for externally imposed transmit-power constraints. To solve the resulting non-convex mixed-integer problem, we develop an alternating-optimization framework that exploits structural properties of the EE function and converges to a stationary point. Simulation results show median EE improvements of up to $3\times$ in large multi-user scenario compared to fixed power and antenna allocation baselines. 
\end{abstract}


\IEEEpeerreviewmaketitle

\section{Introduction}
\label{sec:intro}
\footnotetext[0]{\thanks{Parts of this work have been accepted for publication in \cite{Marwaha2026ICC_EE_PA}. 
}}

Enabling massive multiple access and high data rates, massive multiple input multiple output (M-MIMO) has emerged as a cornerstone technology in $5$G networks and beyond~\cite{bjornson2014massive}. However, future wireless systems are increasingly constrained not only by spectrum and latency but also by energy availability. As network capacity, intelligence, and connectivity scale, energy has become a primary limiting resource rather than a secondary performance metric. Consequently, sustainable growth in wireless networks requires meeting increasing data-rate demands without a proportional increase in energy consumption, placing energy efficiency (EE) as a key design constraint.

This challenge is particularly pronounced in large-scale multi-antenna systems, where hardware power consumption scales with the number of active components, including power amplifiers (PAs), transceivers, and cables. These components can account for up to $65\%$ of the total energy consumption of the base station (BS)~\cite{9678321}. Therefore, classical resource-allocation strategies that focus primarily on spectral efficiency are no longer sufficient. Instead, jointly optimizing system-level parameters, such as transmit power and the number of active antennas, considering realistic PA models is essential for maximizing the bit-per-joule EE of M-MIMO deployments.
Multiple works have proposed energy-efficient resource allocation schemes for M-MIMO systems; comprehensive surveys can be found in~\cite{9678321,8014295,7523234}. However, most existing works neglect the impact of hardware-induced nonlinearities, particularly those stemming from PA imperfections, e.g.,~\cite{10341276,6951974,8094316}.

Operating PAs in their nonlinear region can increase the desired signal power at the cost of nonlinear distortion and increased power consumption~\cite{kryszkiewicz2023efficiency}. This introduces an additional degree of freedom in EE system design. Accordingly, several works incorporate hardware impairments into EE analysis~\cite{6854179,4533658,7031971,bjornson2014massive,9226127,Fettweis_EE_MIMO_EUCNC2025}. However, these studies remain limited in two important aspects. First, most assume narrowband transmission with flat-fading channels and therefore do not capture subcarriers intermodulation in orthogonal frequency division multiplexing (OFDM) systems. Second, hardware impairments are typically modeled either as additive Gaussian distortion of fixed power or as distortion power that grows linearly with the desired signal power~\cite{6854179,7031971,bjornson2014massive}.

While these assumptions simplify analysis, they provide only a local approximation of PA behavior. In particular, modeling distortion power as linearly proportional to the desired signal power results in a constant signal-to-distortion ratio (SDR), which is valid only over a limited transmit-power range, as noted in~\cite{bjornson2014massive}. Consequently, PA operation cannot be optimized across its full operating range, especially near the saturation (clipping) region that is often relevant for EE optimization. As a result, such models typically require externally imposed transmit-power constraints, whereas wide-range-valid PA models naturally bound the feasible operating region through the PA saturation power.

A recent study employed a soft-limiter PA model in a hybrid MIMO–OFDM system but did not perform EE optimization~\cite{Fettweis_EE_MIMO_EUCNC2025}. Related work has also considered refined nonlinear distortion modeling, such as distortion-correlation-aware power allocation in M-MIMO systems~\cite{salman2025distcorr}. However, these works either focus on spectral-efficiency optimization or provide parametric system analyses without solving an EE maximization problem.

Table~\ref{tab:relatedworks} summarizes the assumptions, hardware models, and optimization variables of the most relevant prior works and highlights the modeling gap addressed in this paper.

Despite these advances, the joint EE optimization of transmit power, antenna activation, and per-user power allocation for wideband M-MIMO OFDM systems under realistic nonlinear PA behavior remains largely unexplored.
Motivated by the potential of PA back-off optimization for improving EE,also acknowledged by 3GPP for 5G systems~\cite{3GPP38864}, we aim to maximize the EE of an M-MIMO OFDM BS.

This work formulates and solves an EE-maximization problem for a downlink (DL) multi-user (MU) M-MIMO OFDM BS operating under nonlinear PA distortion and realistic PA power-consumption modeling (extending our single-UE study in \cite{Marwaha2026ICC_EE_PA}). In contrast to approaches in Table~\ref{tab:relatedworks}, which assume PA power consumption to scale linearly with transmit power (i.e., constant efficiency), the adopted model captures the dependence of both distortion and consumed power on the input back-off (IBO), leading to a tighter coupling between rate and power consumption.
Unlike our prior study on distortion-aware sum-rate maximization~\cite{marwaha2025optimaldistortionawaremultiuserpower}, the present problem involves a nonconvex fractional EE objective whose numerator and denominator depend on both transmit power and the number of active antennas through the nonlinear PA characteristics and BS power-consumption model. Optimizing over the antenna dimension introduces a mixed-integer decision space, while the coupling between nonlinear distortion and power consumption leads to fundamentally different stationary-point behavior, requiring a new algorithmic framework.

The main contributions of this work are summarized as follows:
\begin{itemize}

\item We formulate an EE-maximization problem for a DL MU M-MIMO OFDM system employing ZF precoding under independent, identically distributed (i.i.d.) Rayleigh fading channels. A soft-limiter PA model is adopted, yielding analytically tractable SNDR expressions that capture the nonlinear relationship between transmit power, distortion, and PA power consumption. Unlike polynomial PA models that are accurate only within limited power regimes, the soft-limiter model inherently covers the feasible power region. The resulting formulation jointly considers the total transmit power, its distribution across UEs, and the number of active antennas, leading to a mixed-integer nonconvex fractional optimization problem.

\item We develop an alternating-optimization (AO) framework tailored to the proposed EE formulation. The problem decomposes into three blocks: two single-variable nonconvex sub-problems corresponding to transmit power and antenna count, and one multi-variable concave sub-problem corresponding to per-UE power fractions. By analyzing the asymptotic behavior of the EE derivatives, we establish the existence of stationary points for the single-variable sub-problems, enabling efficient numerical solutions via bisection-based root finding. The per-UE power-allocation sub-problem admits a unique global solution via a water-filling structure.

\item We analyze the computational complexity of each AO sub-problem using Bachmann–Landau notation and derive the overall algorithmic complexity. The resulting complexity scales linearly with the number of UEs and logarithmically with the desired numerical precision. Simulation results confirm fast empirical convergence, with the AO algorithm typically stabilizing within a few outer iterations.

\item Numerical results provide several design insights unique to EE maximization under nonlinear PA distortion. In particular, the EE-optimal operating point occurs at finite transmit power and a finite number of active antennas, beyond which additional resources reduce EE due to PA saturation and circuit-power scaling. Furthermore, the optimal per-UE power allocation depends strongly on UEs pathloss disparities, highlighting the need for joint optimization of transmit power, UE power fractions, and antenna activation in distortion-aware M-MIMO systems.

\end{itemize}

The rest of this manuscript is organized as follows: Sec. \ref{sec:sys_mod} describes the system model, providing analytically tractable SNDR and achievable rate expressions under the non-linear PA, along with the total power consumption model of the BS and  the EE optimization problem under consideration. Next, our proposed solution is provided in Sec. \ref{sec:sol}. Finally, the results, conclusion, and future work are presented in Sec. \ref{sec:res} and \ref{sec:conc}, respectively.   

\begin{table*}
\centering
\caption{Comparison with Related Works}
\label{tab:relatedworks}

\setlength{\tabcolsep}{2pt}
\footnotesize
\renewcommand{\arraystretch}{1.0}

\newcolumntype{C}[1]{>{\centering\arraybackslash}p{#1}}
\newcolumntype{L}[1]{>{\raggedright\arraybackslash}p{#1}}

\begin{tabular}{|L{1.8cm}|L{2.35cm}|L{2.6cm}|L{1.95cm}|L{2.55cm}|L{3.8cm}|}
\hline
\textbf{Work} &
\textbf{System} &
\textbf{PA / HW model} &
\textbf{Objective} &
\textbf{Variables} &
\textbf{Distortion / structure} \\ \hline

\cite{10341276,6951974,8094316} &
massive MIMO (mostly narrowband / system-level) &
Ideal/linear PA &
EE (GEE) &
Total power and/or antenna adaptation  &
No nonlinear PA distortion in rate; design driven by circuit/load models \\ \hline

\cite{6854179,4533658,7031971,bjornson2014massive} &
narrowband MU-MIMO (and OFDMA in \cite{4533658}) &
Additive hardware impairment, constant SDR type approx. (EVM-like) locally valid &
EE &
Primarily total power (sometimes antenna scaling), typically no UE-split under nonlinear PA &
Distortion power fixed or linear/proportional; constant SDR; amenable to fractional programming/ convexification \\ \hline

\cite{9226127} &
narrowband MIMO &
Simplified impairment model (constant SDR)  &
EE &
Total power only; UE-level split typically fixed &
Implicit/convexified structure enabled by constant-SDR modeling  \\ \hline

\cite{salman2025distcorr} &
DL MU-MIMO OFDM &
Nonlinear PA (polynomial)  &
SE &
Per-UE power fractions under sum constraint  &
Correlation-aware distortion modeling; not an EE formulation  \\ \hline

\cite{Fettweis_EE_MIMO_EUCNC2025} &
Hybrid MIMO--OFDM &
Soft-limiter (wide-range) + quantization &
only SE--EE characterization &
None (parametric study over back-off / resolution) &
Accurate wide-range PA model; highlights SE--EE sensitivity to PA back-off, but no optimization \\ \hline

\textbf{This Work} &
\textbf{DL MU--MIMO OFDM} &
\textbf{Soft-limiter (wide-range, power-dependent distortion)} &
\textbf{EE} &
\textbf{Total power $P$ + UE power split $\boldsymbol{\omega}$ + antenna count $M$} &
\textbf{Exact nonlinear distortion enters SNDR and EE; mixed discrete-continuous fractional problem solved  via AO} \\ \hline

\end{tabular}
\end{table*}

\section{System Model}
\label{sec:sys_mod}
\begin{figure}
    \centering
    \includegraphics[width=1.0\columnwidth]{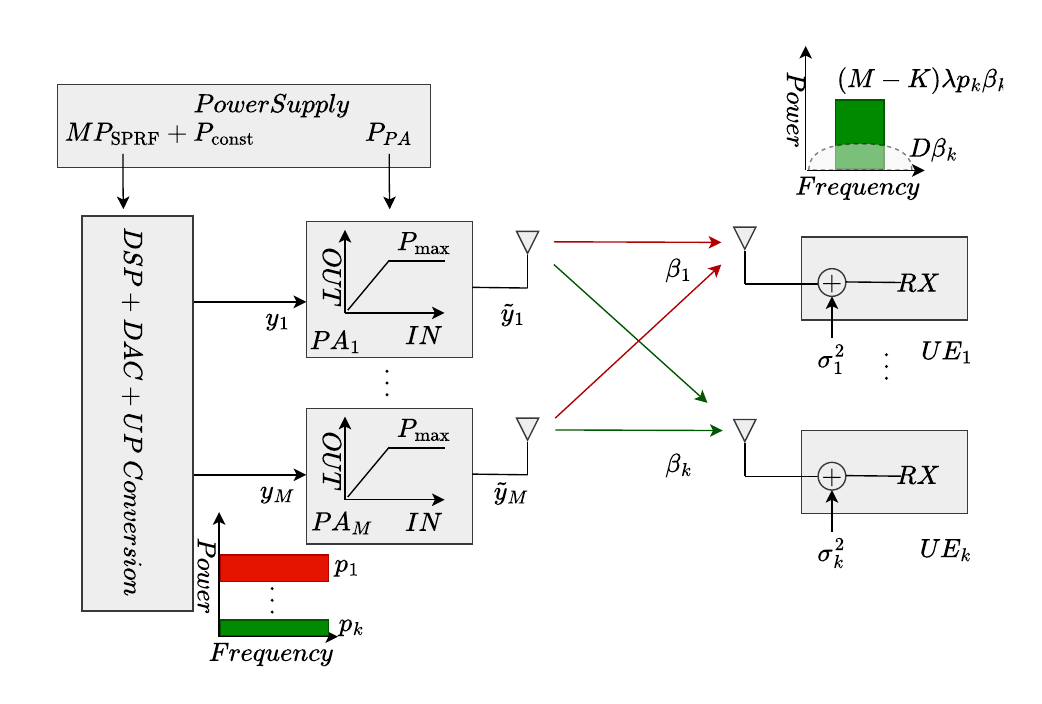}
    \caption{System model: the BS is equipped with $M$ transmit chains, each consisting of common digital signal processing (DSP), digital-to-analog converter (DAC) and up conversion. Using  ZF precoding, the system simultaneously serves $K$ UEs, where the $k$-th UE is characterized by $\beta_k$ mean channel gain and total (over all antennas) allocated power $p_k$.} 
    \label{fig:model}
\end{figure}
As depicted in Fig.~\ref{fig:model}, we consider a DL MU M-MIMO OFDM system, where an $M$-antenna BS simultaneously serves $K$ single-antenna UEs, reusing the setup from \cite{marwaha2025optimaldistortionawaremultiuserpower}. The wireless channels between the BS antennas and the UEs follow an i.i.d. Rayleigh fading model, while the large-scale fading coefficient of the $k$-th UE is denoted by $\beta_k$.

To mitigate inter-UE interference, the BS employs ZF precoding assuming $K<M$, enabling simultaneous transmission to all UEs across the available subcarriers \cite{massivemimobook}. The system utilizes an OFDM waveform with $N_{\mathrm{U}}\gg 1$ subcarriers transmitted in parallel. Under this condition, the time-domain signal at each PA input can be accurately modeled as complex Gaussian distributed due to the central limit theorem \cite{Wei_2010_dist_OFDM}. For detailed signal processing steps we refer the reader to \cite{Wachowiak2023}.

Let $p_k$ denote the transmit power allocated to the $k$-th UE. The total transmit power equals
\begin{equation}
P=\sum_k p_k,
\end{equation}
resulting in a mean input power of $P/M$ at each PA.

Each transmit chain includes a nonlinear PA. We consider a soft-limiter PA model that linearly amplifies the signal when the instantaneous power is below the saturation power $P_{\mathrm{max}}$ and clips it otherwise. The operating point of the PA is characterized by the IBO
\begin{equation}
\Psi =\frac{P_{\mathrm{max}}}{P/M}.
\label{eq_IBO_def}
\end{equation}

Using Bussgang decomposition, the PA output can be expressed as a scaled desired signal plus an uncorrelated distortion component. The signal scaling factor $\lambda\in(0,1)$ depends only on the IBO and equals
\begin{equation}
\lambda =
\left(
1-e^{-\Psi}
+
\frac{1}{2}\sqrt{\pi \Psi}\,
\mathrm{\textrm{erfc}}\left(\sqrt{\Psi}\right)
\right)^2,
\label{eq_lambda}
\end{equation}
while the effective distortion power aggregated over all utilized subcarriers is
\begin{equation}
D= \frac{2}{3}\left(1-e^{-\Psi}-\lambda \right)P.
\end{equation}

Assuming that nonlinear distortion from different transmit chains combines incoherently at the receiver and treating it as noise-like interference added to white noise of total power $\sigma_k^2$ over allocated subcarriers, the resulting signal-to-noise-and-distortion ratio (SNDR) for the $k$-th UE under ZF precoding is \cite{marwaha2025optimaldistortionawaremultiuserpower}
\begin{equation}
\gamma_k =  \frac{(M-K) \lambda p_k \beta_k}{\sigma_k^2 + \beta_k D }.
\label{eq: SINR_ZF}
\end{equation}
The achievable throughput of the $k$-th UE is therefore
\begin{equation}
R_k = N_{\mathrm{U}} \Delta f \log_2 (1 + \gamma_k),
\label{eq: datarate2}
\end{equation}
where $\Delta f$ is the OFDM subcarrier spacing. Although nonlinear distortion can increase channel capacity when advanced receivers are employed \cite{Guerreiro_2015_optimum_performance}, such techniques incur significantly higher receiver complexity and are not considered here.

The achievable rate in \eqref{eq: datarate2} is obtained at the cost of the consumed transmitter power. The total BS power consumption is modeled as \cite{10341276}
\begin{equation}
P_{\mathrm{tot}} = P_{\mathrm{PA}}+P_{const} + M P_{\mathrm{SPRF}},
\label{eq: egy_cons2}
\end{equation}
where $P_{\mathrm{PA}}$ denotes the power consumed by the PAs, $P_{\mathrm{const}}$ accounts for static power consumption (e.g., site cooling and local oscillators), and $P_{\mathrm{SPRF}}$ represents the per-antenna RF chain power consumption (e.g., mixers).

The PA power consumption depends on the PA architecture and the instantaneous output signal distribution $|\tilde{y}_{m}|^2$ (denoted in Fig. \ref{fig:model}) \cite{ochiai2013analysis}. For OFDM transmission with a soft-limiter PA model, the total consumed PA power across all antennas equals \cite{kryszkiewicz2023efficiency,ochiai2013analysis}
\begin{equation}
P_{\mathrm{PA}} =
\begin{cases}
\frac{2MP_{\mathrm{max}}}{\sqrt{\pi \Psi}}\mathrm{erf}(\sqrt{\Psi}), & \textbf{Class B PA} \\
\frac{MP_{\mathrm{max}}}{\Psi} (1- e^{-\Psi}), & \textbf{Perfect PA}
\end{cases}
\label{eq:p_pa}
\end{equation}
where the Perfect PA represents an idealized architecture, consuming exactly the radiated power (i.e., the sum of useful signal and distortion components).

We aim to maximize the EE of the system by jointly optimizing the transmit power allocation $\boldsymbol{p}=[p_1,\ldots,p_K]$ and the number of active antennas $M$. The EE is defined as
\begin{subequations}
\begin{alignat}{2}
\max_{\boldsymbol{p}\geq 0, M \in \mathbb{N}^+}   &\quad  
EE = \frac{\sum_k R_k}{P_{\mathrm{tot}}}
\label{eq:ee}\\
\mathrm{s.t.} &\quad M>K,
\label{eq_problem_constraint1}
\end{alignat}
\label{eq:optprob}
\end{subequations}
where \eqref{eq_problem_constraint1} follows from the ZF precoding requirement.

Problem \eqref{eq:optprob} is a non-convex mixed-integer fractional program. While the power allocation vector $\boldsymbol{p}$ lies in the continuous domain, the antenna count $M$ is an integer variable, resulting in a integer search space. Moreover, both the numerator and denominator of \eqref{eq:ee} are non-convex due to the nonlinear PA model, which complicates even the continuous relaxation of the problem.

Although fractional programming techniques such as Dinkelbach's method can be used to solve EE optimization problems, the non-convexity of the objective function prevents its direct application. Hybrid approaches combining Dinkelbach's method with successive convex approximation (SCA) \cite{zappone2014energy} are possible but require nested iterative procedures that increase computational complexity. Instead, we develop a lower-complexity solution tailored to the structure of the EE function in the next section.

\section {EE Optimization Algorithm}
\label{sec:sol}
To solve \eqref{eq:optprob}, we adopt an AO framework by decoupling the joint optimization problem over $\boldsymbol{p}$ and $M$ into more tractable optimization sub-problems, i.e., we optimize $\boldsymbol{p}$ for a fixed $M$ by solving
\begin{align}
\max_{\boldsymbol{p}\geq0}   &\quad  EE = \frac{\sum_k R_k}{P_{tot}}.
\label{eq:optprob1}
\end{align}
We call this Distortion-aware Energy Efficient Power allocation ({DEEP}). 
Next, we optimize $M$ for a fixed $\boldsymbol{p}$ solving 
\begin{align}
\max_{M>K}   &\quad  EE = \frac{\sum_k R_k}{P_{tot}},
\label{eq:optprob2}
\end{align}
which we call Distortion-aware Energy efficient Antenna aLlocation ({DEAL}). 

Observe that \eqref{eq: datarate2} is non-convex. Therefore, we decouple the first sub-problem in \eqref{eq:optprob1} further using an AO framework, which is discussed in Sec. \ref{subsec:opt_P_omega}. In this framework, we first solve the distortion-aware total power allocation problem, followed by the distribution of the fixed allocated power among the UEs. 
Utilizing the properties of the EE function, we prove that bisection with function-specific starting point search can be employed to find the maximum of the distortion-aware total power allocation problem. On the other hand, although the power distribution among UEs is a multi-variable optimization problem, it appears to be concave, and the solution exhibits a water-filling structure.  

Exploiting the structure of EE function with respect to the relaxed $M$, i.e., sub-problem \eqref{eq:optprob2}, we show in Sec. \ref{subsec:opt_M} that its derivative with respect to $M$ is continuous and has a unique root, which is efficiently found via a bisection-based root finding method with function-specific starting point search. Thus, we avoid an extra step of employing SCA, otherwise required with Dinkelbach's framework, guarantee convergence to a stationary point, and achieve optimal performance at significantly reduced computational cost.

\subsection{Distortion-Aware Energy Efficient Power Allocation (DEEP)}
\label{subsec:opt_P_omega}
The optimization problem in \eqref{eq:optprob1} is non-convex, which can be justified just by the non-convexity of \eqref{eq: datarate2}, as shown formally in \cite{marwaha2025optimaldistortionawaremultiuserpower}, as a result of the nonlinear PA model. Therefore, solving for individual power allocation $p_k$ is non-trivial. Thus, we decouple it as $p_k = \omega_k P$, where $P$ represents the total allocated power across all PAs supporting data transmission to all $K$ UEs and $\omega_k$ denotes the fraction of $P$ allocated to UE $k$, requiring that $\omega_k\geq 0$ and $\sum_k \omega_k =1$ as $\sum_k p_k=P \sum_k \omega_k =P$. This allows to deconstruct \eqref{eq:optprob1} in two sub-problems and adopt an AO framework, enabling us to first find the optimum distortion-aware total power allocation by solving  
\begin{align}
\max_{P\geq0}   &\quad  EE = \frac{\sum_k R_k}{P_{tot}},
\label{eq:optprob1_P}
\end{align}
followed by the distribution of the distortion-aware fixed power allocation algorithm finding the optimum ($\boldsymbol{\omega} = [\omega_1,\dots,\omega_K]$) by solving  
\begin{subequations}
    \begin{alignat}{2}
    \max_{\boldsymbol{\omega} \geq \boldsymbol{0}} &\quad& & EE = \frac{\sum_k R_k}{P_{\mathrm{tot}}} \label{eq:omega_optprob1} \\
    \text{s.t.} &\quad& & \sum_k \omega_{k} = 1.\label{eq:sum_const} 
    \end{alignat}
    \label{eq:optprob1_omega}
\end{subequations}
The solution to the non-convex distortion-aware total power allocation problem equates to finding the stationary point of \eqref{eq:optprob1_P}, which can be achieved by solving 
\begin{equation}    
\at{\frac{\partial {EE}}{\partial P}}{P=\tilde{P}}=0
    \label{eq_EE_d_Rk}
\end{equation}
for the root $\tilde{P}$. In order to find $\tilde{P}$, first properties of $\frac{\partial {EE}}{\partial P}$ can be analyzed using the asymptotic expansion of the functions therein. This enables us to propose the following lemma: 
\begin{lemma}\label{lem:root_find}
For a fixed $M$ and $\boldsymbol{\omega}$, the function $\frac{\partial {EE}}{\partial P}$ has at least a single root equal to the root of the function
\begin{equation}
f_P(P)= \frac{1}{\sum_k R_k}\frac{\partial {\sum_k R_k}}{\partial P} - \frac{1}{P_{\mathrm{tot}}}\frac{\partial P_{\mathrm{tot}}}{\partial P}
\label{eq_f}
\end{equation}
in the entire range of $P$, i.e., $P \in [ 0, \infty )$. 
\end{lemma}
\begin{proof}
The proof of Lemma \ref{lem:root_find} is provided in Appendix \ref{sec:proof_lemma_1}, where the derivative $\frac{\partial {EE}}{\partial P}$ is computed and its characteristic is analyzed. Our analysis shows that $f_P(P) \to +\infty$ as $P \to 0$, whereas, $f_P(P)\to 0^-$ as $P\to+\infty$. Thus, by Intermediate Value Theorem, there must exist at least one root. 
\end{proof}
Lemma \ref{lem:root_find} enables us to propose a bisection-based Algorithm~\ref{alg:generic_bisection}, instantiated for {DEEP} by setting: $x_{\text{test}} \gets P_{\text{test}}$, $x_L \gets P_L$, $x_U \gets P_U$, $x_C \gets P_C$, $f_x(x) \gets f_P(P)$, and $\delta_x \gets \delta_P$, to find the root of \eqref{eq_f}.
The asymptotic behavior of $f_P(P)$ ensures that for sufficiently low $P_{L}$, the derivative $f_P(P_{L})$ is greater than 0. Similarly, for sufficiently high $P_{U}$, the derivative $f_P(P_{U})$ is smaller than 0. Two initial \emph{while} loops aim at finding such points. Next, the bisection search method is used to find the root $P_C \in [P_{L}, P_{U}]$, which is limited by a non-negative value $\delta_P$, i.e., the root belongs to the range $[P_{C}-\frac{\delta_P}{2}, P_{C}+\frac{\delta_P}{2}]$.     
\begin{algorithm}
\caption{Bisection-Based Root-Finding Algorithm with function-specific starting point search}
\label{alg:generic_bisection}
\begin{algorithmic}[1]
\State Initialize: $x_{\text{test}}$, $x_L \gets x_{\text{test}}$, $x_U \gets x_{\text{test}}$, $\delta_x$
    \While{$\left.f_x(x)\right|_{x=x_U} > 0$} 
        \State $x_L \gets x_U$, \; $x_U \gets 2x_U$
    \EndWhile
    \While{$\left.f_x(x)\right|_{x=x_L} < 0$}
        \State $x_U \gets x_L$, \; $x_L \gets 0.5x_L$
    \EndWhile
    \While{$|x_U - x_L| > \delta_x$}
    \State $x_C \gets 0.5x_L + 0.5x_U$
    \If{$\left.f_x(x)\right|_{x=x_C} > 0$} 
        \State $x_L \gets x_C$
    \Else
        \State $x_U \gets x_C$
    \EndIf
\EndWhile
\State $x_C \gets 0.5x_L + 0.5x_U$
\end{algorithmic}
\end{algorithm}

After finding $P$, the optimum distribution of the allocated total power $P$ among the UEs can be achieved by applying the Karush-Kuhn-Tucker (KKT) conditions, enabling us to propose the following lemma:
\begin{lemma} \label{lem:P_dist}
    For a fixed $M$ and $P$, the optimal solution of the optimization problem in \eqref{eq:optprob1_omega}, represented by $\omega^*_k$, is
    \begin{align}
        \omega^*_k=\max \left\{ 0 , \frac{N_{\mathrm{U}} \Delta f}{\nu^*P_{\mathrm{tot}}} - \frac{\sigma^2_k + \beta_kD}{(M-K)\lambda P\beta_k} \right\},
    \label{eq:opt_omega}
    \end{align}
    where $\nu^*$ is the optimal solution of the Lagrange dual problem, which is found by ensuring that $\sum_k\omega^*_k = 1$. 
\end{lemma}
\begin{proof}
    Observe that by substituting the expressions for $R_k$ from \eqref{eq: datarate2} and $\gamma_k$ from (\ref{eq: SINR_ZF}) into the objective function of \eqref{eq:optprob1_omega} following the substitution $p_k = P\omega_k$, the optimization problem simplifies to
    \begin{align}
        \max_{\{\omega_k\}} \quad \sum_k \tilde{B}\log_2\!\left(1 + A_k \, \omega_k\right),
    \end{align}
    where $\tilde{B} \!=\! \frac{N_{\mathrm{U}} \Delta f}{P_{\mathrm{tot}}}$ and the UE-specific constant $A_k \!=\! \frac{(M-K)\lambda P\beta_k}{\sigma_k^2 + \beta_k D}$, representing the standard water-filling problem \cite{Boyd_Vandenberghe_2004}. By formulating the Lagrangian and applying the KKT conditions, the optimal distortion-aware power distribution can be obtained by the expression in \eqref{eq:opt_omega}.
\end{proof}
The solution to $\omega_k^*$ in \eqref{eq:opt_omega} resembles the well known water-filling solution, which is a piecewise-linear increasing function of $\frac{N_U \Delta f}{\nu^*P_{\mathrm{tot}}}$, with break points at $\frac{\sigma^2_k + \beta_kD}{(M-K)\lambda P\beta_k}$ for each $k$. Therefore, the optimum $\nu^*$ can be found either by the well-known bisection method \cite{Boyd_Vandenberghe_2004} or optimized solutions, e.g., \cite{8995606}.

\subsection{Distortion-Aware Energy Efficient Antenna Allocation (DEAL)}
\label{subsec:opt_M}
While the number of antennas $M$ is an integer number, we employ integer relaxation, assuming initially that $M$ is a positive, real number.
To optimize the number of antennas, we propose identifying a stationary point by solving 
\begin{equation}    
\at{\frac{\partial {EE}}{\partial M}}{M=\tilde{M}}=0,
    \label{eq_EE_d_R_M}
\end{equation}
for the root $\tilde{M}$. In order to find $\tilde{M}$, the properties of $\frac{\partial {EE}}{\partial M}$ can be analyzed, resulting in the following lemma: 
\begin{lemma}\label{lem:root_find_M}
For a fixed $P$ and $\boldsymbol{\omega}$, the function $\frac{\partial {EE}}{\partial M}$ has at least a single root equal to the root of the function
\begin{equation}
f_M(M)= \frac{1}{\sum_k R_k}\frac{\partial {\sum_k R_k}}{\partial M} - \frac{1}{P_{\mathrm{tot}}}\frac{\partial P_{\mathrm{tot}}}{\partial M}
\label{eq_f_M}
\end{equation}
in the entire range of $M$, i.e., $M \in [ K^+, \infty )$. 
\end{lemma}
\begin{proof}
By relaxing $M$ to the continuous domain, the derivative of the EE function can be expressed as
\begin{equation}
\frac{\partial \mathrm{EE}}{\partial M} =
\frac{\left(\sum_k \frac{\partial R_k}{\partial M}\right) P_{\mathrm{tot}} - \left(\sum_k R_k\right)\frac{\partial P_{\mathrm{tot}}}{\partial M}}{P_{\mathrm{tot}}^2}.
\end{equation}
Since $P_{\mathrm{tot}} > 0$, the stationary points are determined by the numerator, which leads to the function $f_M(M)$ in \eqref{eq_f_M}, which is continuous for $M > K$.
The derivatives $\partial R_k/\partial M$ and $\partial P_{\mathrm{tot}}/\partial M$ follow directly from the definitions of $\gamma_k$ and $P_{\mathrm{tot}}$. Analyzing the asymptotic behavior, similarly as was done for $f_P(P)$ in Appendix \ref{sec:proof_lemma_1}, we observe that $f_M(M) \to +\infty$ as $M \to K^+$, which follows from the vanishing SNDR as $(M-K)\to 0^+$. Whereas, $f_M(M)\to 0^-$ as $M\to+\infty$ because the achievable rate grows at most logarithmically with $M$, while the total power consumption increases at least linearly due to RF-chain and PA contributions. Therefore, by the Intermediate Value Theorem, there exists at least one root of $f_M(M)$ in the interval $[K^+, \infty)$.
\end{proof}

Lemma \ref{lem:root_find_M} enables us to use bisection-based Algorithm \ref{alg:generic_bisection}, instantiated for {DEAL} by setting: $x_{\text{test}} \gets M_{\text{test}}$, $x_L \gets M_L$, $x_U \gets M_U$, $x_C \gets M_C$, $f_x(x) \gets f_M(M)$, and $\delta_x \gets \delta_M$, to find the root of \eqref{eq_f_M}.
The asymptotic behavior of $f_M(M)$ ensures that for sufficiently low $M_{L}$, the derivative $f_M(M_{L})$ is greater than 0. Similarly, for sufficiently high $M_{U}$, the derivative $f_M(M_{U})$ is smaller than 0. Two initial \emph{while} loops aim at finding such points. Next, the bisection search method is used to find the root $M_C \in [M_{L}, M_{U}]$, which is limited by a non-negative value $\delta_M$, i.e., the root belongs to the range $[M_{C}-\frac{\delta_M}{2}, M_{C}+\frac{\delta_M}{2}]$.   

The obtained continuous root $M_c$ plays a vital role in identifying the optimal antenna count. The following proposition shows how the continuous root can be used to find the optimal integer-valued solution.
\begin{prop} \label{prop1}
    Assume that EE(M) is a unimodal function with respect to M. Let $M_C \in \mathbb{R}$ denote the root of $f_M(M)$ in \eqref{eq_f_M}, i.e.,  $\exists_{M\in [M_C - \delta_M/2, M_C + \delta_M/2] } f_M(M)=0$, obtained using Algorithm \ref{alg:generic_bisection} instantiated with $x \gets M$ and tolerance $\delta_M\ll1$. 
    Then the optimal integer antenna count that maximizes $EE(M)$ in \eqref{eq:optprob2}, i.e.,
    \begin{equation}
        M^* = \arg \max_{M\in \mathbb{Z}, M\geq K+1} EE(M),
    \end{equation}
    belongs to the set $\{\left\lfloor M_C \right\rfloor, \left\lceil M_C \right\rceil\}$.
\end{prop}
\begin{proof}
    Since EE(M) is unimodal, there exists a unique maximizer $M_c \in \mathcal{R}$ of the continuous relaxation. By definition of unimodality, EE(M) is strictly increasing for $M<M_c$ and strictly decreasing for $M>M_c$ Hence, among all integers less than $M_c$, the maximum is attained at $\left\lfloor M_C \right\rfloor$ and among all integers greater than $M_c$, the maximum is attained at $\left\lceil M_C \right\rceil$. Therefore, the global optimal integer solution must belong to the set $\{\left\lfloor M_C \right\rfloor, \left\lceil M_C \right\rceil\}$
\end{proof}

\begin{remark}
\begin{figure}[ht]
    \centering
    \includegraphics[width=0.8\columnwidth]{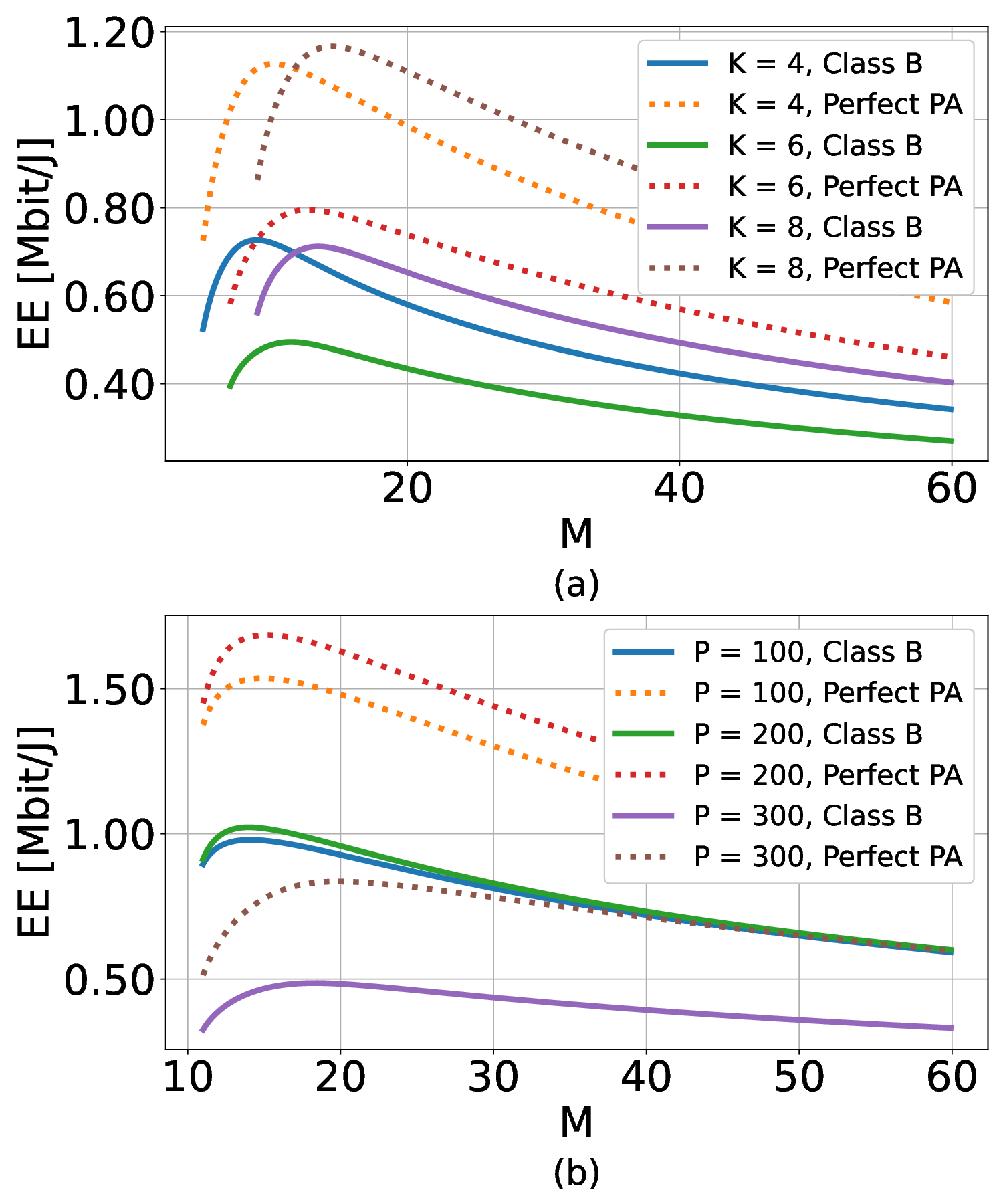}
    \caption{Plots of EE(M) for system parameters as used in Sec. \ref{sec:res} (see Table \ref{tab: sys_para})
    with fixed $P=200$ W and $K=10$ for figure a) and b), respectively. For each series, $\beta_k$ and $\omega_k$ is selected randomly from uniform distribution over $(60,150)$ dB, and a Dirichlet distribution with $\omega_k \geq 0$ and $\sum_k \omega_k = 1$, respectively. 
   } 
    \label{fig:unimodal_EE}
\end{figure}
Observe that the function $EE(M)$ is non-convex due to the non-convexity of \eqref{eq: datarate2}, as well as due to the fractional form of the objective function. Although it has been shown via Lemma \ref{lem:root_find_M} that $f_M(M) \to +\infty$ as $M \to K^+$, whereas $f_M(M)\to 0^-$ as $M\to+\infty$, the behavior between these extreme points cannot be determined in closed-form. Despite this, extensive numerical evidence under diverse representative system parameters (Fig. \ref{fig:unimodal_EE}) consistently shows that $EE(M)$ monotonically increases up to $M_c$ and decreases thereafter, strongly suggesting a single maximum and supporting the use of Proposition \ref{prop1} for selecting $M^*$. 
\end{remark}

\subsection{Alternating Optimization}
\label{subsec:AO}

The solutions of sub-problems presented in Sec. \ref{subsec:opt_P_omega} and Sec. \ref{subsec:opt_M} can be used to solve the optimization problem \eqref{eq:optprob1} and \eqref{eq:optprob2}, and thereby \eqref{eq:optprob}, by AO as shown in Algorithm \ref{alg:ao}. Firstly, for a fixed $M$ and $\boldsymbol{\omega} = [\omega_1,\dots, \omega_k]$, we find the optimum solution for \eqref{eq:optprob1_P} via Algorithm \ref{alg:generic_bisection} instantiated with $x \gets P$. Then for the fixed M and with this optimum $P$, we find the optimum solution for \eqref{eq:optprob1_omega} via well-known bisection method with accuracy $\delta_{\nu}$ \cite{Boyd_Vandenberghe_2004}. 
Finally, for the optimum $P$ and $\boldsymbol{\omega}$, we find the optimum $M$, i.e. solution to \eqref{eq:optprob2} via Algorithm \ref{alg:generic_bisection} instantiated with $x \gets M$. The process is repeated over multiple iterations of $P$ and $\boldsymbol{\omega}$, and $M$\textemdash with superscript $(i)$ denoting the solution at the $i$-th iteration\textemdash until convergence is achieved, i.e, until the difference between EE in the current and previous iteration is smaller than a non-negative value $\delta_{EE}$, which is the non-negative convergence tolerance. Finally, after finding solutions over the continuous domain, we evaluate EE at $\left\lfloor M \right\rfloor$ and $\left\lceil M \right\rceil$ and choose the one that results in the highest EE.
\begin{algorithm}
\caption{Alternating Optimization Framework for EE Maximization}\label{alg:ao}
\begin{algorithmic}[1]
\State Initialize: $i \gets 0$,  $\boldsymbol{\omega}^{(i)} = [\omega_1^{(i)}, \dots, \omega_K^{(i)}] \gets \frac{1}{K}$, $P \gets P^{(i)}$, $M \gets M^{(i)}$, $\delta_{EE}$, $EE^{(0)} \gets \mathrm{EE}(P^{(0)},\boldsymbol{\omega}^{(0)}, M^{(0)})$
\Repeat
\State $i \gets i+1$
\State solve \eqref{eq:optprob1_P} given $M^{(i-1)}$ and $\boldsymbol{\omega^{(i-1)}}$ to get $P^{(i)}$ by \textbf{ Algorithm \ref{alg:generic_bisection}} for $f_P(P)$
\State solve \eqref{eq:optprob1_omega} given $M^{(i-1)}$ and $P^{(i)}$ to get $\boldsymbol{\omega^{(i)}}$ by bisection method \cite{Boyd_Vandenberghe_2004}
\State solve \eqref{eq:optprob2} given $P^{(i)}$ and $\boldsymbol{\omega^{(i)}}$ to get $M^{(i)}$ by \textbf{ Algorithm \ref{alg:generic_bisection}} for $f_M(M)$
\State Compute: $\mathrm{EE}(P^{(i)},\boldsymbol{\omega}^{(i)}, M^{(i)})$
\Until${|\mathrm{EE}^{(i)} - \mathrm{EE}^{(i-1)}|< \delta_{EE}}$
\If{$\mathrm{EE}(P^{(i)},\boldsymbol{\omega}^{(i)}, \lfloor M^{(i)}\rfloor)>\mathrm{EE}(P^{(i)},\boldsymbol{\omega}^{(i)}, \lceil M^{(i)}\rceil)$}
\State $M^{*}\gets \left\lfloor M^{(i)}\right\rfloor$
\Else
\State $M^{*}\gets \left\lceil M^{(i)}\right\rceil$
\EndIf
\end{algorithmic}
\end{algorithm}

\subsection{Convergence Properties of Algorithm \ref{alg:ao}}
\label{subsec:convergence}

The proposed DEEP--DEAL framework employs a cyclic alternating-optimization scheme over three variable blocks: the total transmit power $P$, the power-fraction vector $\boldsymbol{\omega}$, and the (relaxed) number of active antennas $M$. We establish convergence in the sense of block-coordinate stationarity by verifying the conditions of classical block-coordinate optimization results, i.e., Theorem 4.1 in \cite{tseng2001convergence}.

\begin{lemma}
Consider the relaxed EE maximization problem
\begin{align}
\max_{P\ge0,\;\boldsymbol{\omega}\in\Delta_K,\;M> K}
\mathrm{EE}(P,\boldsymbol{\omega},M),
\end{align}
where $\Delta_K \triangleq \{\boldsymbol{\omega}\succeq \mathbf{0} :
\mathbf{1}^\top \boldsymbol{\omega}=1\}$ and let $\mathbf{x}^{(i)}=(P^{(i)},\boldsymbol{\omega}^{(i)},M^{(i)})$ denote
the variables after the $i$-th AO cycle of the DEEP-DEAL algorithm with objective value $\mathrm{EE}(\mathbf{x}^{(i)})$,
which cyclically maximizes the blocks in the order
$P \rightarrow \boldsymbol{\omega} \rightarrow M$.
If each block subproblem is solved exactly and $\mathrm{EE}$ is continuous over the feasible set, then the sequence $\{\mathrm{EE}(\mathbf{x}^{(i)})\}$ is monotonically non-decreasing and the
iterates remain in the upper level set
\begin{align}
\mathcal{L}=
\{(P,\boldsymbol{\omega},M):
\mathrm{EE}(P,\boldsymbol{\omega},M)
\ge
\mathrm{EE}(\mathbf{x}^{(0)})\}.
\end{align}
If this level set is closed and bounded, then $\mathcal{L}$ is compact and every cluster point of the sequence generated by DEEP-DEAL is a coordinatewise optimal point of the relaxed problem. Furthermore, if $\mathrm{EE}$ is continuously differentiable and the constraints are affine, every such cluster point satisfies the first-order KKT conditions.
\end{lemma}

\begin{proof}
\emph{1) Monotonic ascent.}
Lemma \ref{lem:root_find}, \ref{lem:P_dist}, and \ref{lem:root_find_M} ensure that each block update maximizes $\mathrm{EE}$ while keeping the remaining variables fixed. Hence
\begin{align}
\mathrm{EE}(\mathbf{x}^{(i+1)})
\ge
\mathrm{EE}(\mathbf{x}^{(i)}), \quad \forall i,
\end{align}
and the EE sequence is monotonically non-decreasing. Consequently, all iterates remain in the upper level set $\mathcal{L}$.

\emph{2) Compactness of the upper level set.}
Firstly, The simplex $\Delta_K$ is compact. Next, in the high–power regime $P \to \infty$, the SNDR converges to a finite constant \eqref{eq:gamma_approx}, implying that the achievable rate remains bounded. Moreover, the PA power saturates yielding
\begin{equation}
\lim_{P\to\infty} P_{\mathrm{PA}}=
\begin{cases}
\frac{4}{\pi} M P_{\max}, & \textbf{Class B PA},\\
M P_{\max}, & \textbf{Perfect PA},
\end{cases}
\end{equation} 
and therefore, the total consumed power converges to a finite  constant. Consequently,
\begin{equation}
\lim_{P \to \infty}
\mathrm{EE}(P,\boldsymbol{\omega},M)
= \mathrm{EE}_{\infty}(\boldsymbol{\omega},M) < \infty, 
\end{equation}
where 
\begin{equation}
\mathrm{EE}_{\infty}(\boldsymbol{\omega},M)
=
\frac{
N_{\mathrm U}\Delta f
\sum_{k=1}^{K}\log_2(1+\gamma_{k,\infty})
}
{MP_{\max}+MP_{\mathrm{SPRF}}+P_{\mathrm{const}}}.
\end{equation}
In general, the asymptotic limit $\mathrm{EE}_{\infty}(\boldsymbol{\omega},M)$ need not be zero. To ensure that the relevant upper level set generated by the algorithm is bounded, we impose the initialization condition 
\begin{equation}
\mathrm{EE}(\mathbf{x}^{(0)})
>
\sup_{\boldsymbol{\omega}\in\Delta_K,\; M>K}
\mathrm{EE}_{\infty}(\boldsymbol{\omega},M).
\label{eq:init_cond}
\end{equation}
and therefore, $x^{(0)}$ should be selected such that \eqref{eq:init_cond} is satisfied. Such initialization is not restrictive. Since $\mathrm{EE}(P,\boldsymbol{\omega},M)$ converges to $\mathrm{EE}_{\infty}(\boldsymbol{\omega},M)$ as $P \to \infty$ and the derivative $\partial \mathrm{EE}/\partial P \to 0^{-}$ in this regime (Appendix~\ref{sec:proof_lemma_1}), there always exists a finite transmit-power region where the achieved EE exceeds its asymptotic tail value. Therefore an initialization satisfying \eqref{eq:init_cond} can always be selected. 

Assuming the initialization satisfies \eqref{eq:init_cond} and as $P \to \infty$ $\mathrm{EE}(P,\boldsymbol{\omega},M) \to
\mathrm{EE}_{\infty}(\boldsymbol{\omega},M)$, 
it follows from the definition of the limit that there exists a finite constant $P_{c}$ independent of $(\boldsymbol{\omega}, M)$ such that 
\begin{equation}
    \mathrm{EE}(P,\boldsymbol{\omega},M) < \mathrm{EE}(\mathbf{x}^{(0)}) \quad \mathrm{for~all} \quad P > P_{c}.
\end{equation} 
Consequently, points with $P>P_{c}$ cannot belong to the upper level set generated by the algorithm. Therefore, all iterates satisfy $P \in [0,P_{c}]$, which establishes that the upper level set is closed and bounded with respect to $P$.

In the ZF-feasible boundary regime, the SNDR vanishes linearly with $(M-K)$, implying that $\lim_{M \to K^{+}} \mathrm{EE} (P,\boldsymbol{\omega},M)= 0 $.
Since the initialization uses $P^{(0)} > 0$ and $M^{(0)} > K$, the achievable rates are strictly positive, while the total consumed power is strictly positive due to the circuit power terms in the power model. Consequently, $\mathrm{EE}(\mathbf{x}^{(0)}) > 0$.
Since $\mathrm{EE}(\mathbf{x}^{(0)}) > 0$, there exists $M_{\min} > K$ such that
\begin{equation}
\mathrm{EE}(P,\boldsymbol{\omega},M)
<
\mathrm{EE}(\mathbf{x}^{(0)})
\quad
\text{for all } M \in (K, M_{\min}).
\end{equation}
Hence points arbitrarily close to the boundary $M = K$ cannot belong to the upper level set.

In the large–array regime $M \to \infty$, the achievable rate grows at most logarithmically with $M$, whereas the total consumed power increases at least linearly due to RF–chain and PA power contributions. Therefore, 
\begin{equation}
\lim_{M \to \infty}
\mathrm{EE}(P,\boldsymbol{\omega},M)
=
0 .
\end{equation}
Thus there exists a finite constant $M_{\max}$ such that
\begin{equation}
\mathrm{EE}(P,\boldsymbol{\omega},M)
<
\mathrm{EE}(\mathbf{x}^{(0)})
\quad
\text{for all } M > M_{\max}.
\end{equation}

Combining the above results, any point belonging to $\mathcal{L}$ must satisfy
\begin{equation}
P \in [0,P_{c}],
\qquad
M \in [M_{\min},M_{\max}].
\end{equation}
Hence $\mathcal{L}$ is closed and bounded in $(P,M)$ and therefore compact.

\emph{3) Coordinatewise optimality.}
Since the EE function is continuous on the compact level set $\mathcal{L}$ and the algorithm performs exact cyclic block updates, the conditions of \cite[Thm.~4.1]{tseng2001convergence} are satisfied. Therefore every cluster point of the generated sequence is a coordinatewise maximum of $\mathrm{EE}$.

\emph{4) Stationarity.}
Since $\mathrm{EE}$ is continuously differentiable and the constraints are affine, coordinatewise optimality implies first-order KKT stationarity under a standard constraint qualification
\cite[Prop.~2.7.1]{BertsekasNLP3}.
\end{proof}

\subsection{Complexity Analysis of Algorithm \ref{alg:ao}}
\label{subsec:complexity}

We briefly discuss the computational complexity of the proposed DEEP--DEAL framework. Each iteration of the Algorithm \ref{alg:ao} consists of three block updates, i.e., over $P$, $\boldsymbol{\omega}$, and $M$. In all blocks, the dominant cost arises from evaluating the EE and its derivative, which requires computing the SNDR and achievable rate of all $K$ UEs, resulting in a per-evaluation complexity of $\mathcal{O}(K)$.

The $P$-update and the relaxed $M$-update are both solved via bisection with function-specific bracketing. For numerical tolerances $\delta_P$ and $\delta_M$, respectively, each update requires $\mathcal{O}\!\left(\log\!\left(\frac{P^{(i)}}{\delta_P}\right)\right)$ and $\mathcal{O}\!\left(\log\!\left(\frac{M^{(i)}}{\delta_M}\right)\right)$ iterations at the $i$-th AO step, yielding complexities of $\mathcal{O}\!\left(K\log\!\left(\frac{P^{(i)}}{\delta_P}\right)\right)$ and $\mathcal{O}\!\left(K\log\!\left(\frac{M^{(i)}}{\delta_M}\right)\right)$.
For fixed $(P,M)$, the $\boldsymbol{\omega}$-update admits a water-filling-type solution. The associated dual variable is obtained via a one-dimensional bisection search, where each iteration requires computing all $K$ power fractions, resulting in a complexity of $\mathcal{O}\!\left(K\log\!\left(\frac{\nu^{(i)}}{\delta_\nu}\right)\right)$. The final mapping from the relaxed antenna solution to the optimal integer value requires evaluating EE for a small number of candidates and incurs a negligible cost of $\mathcal{O}(K)$.

Combining all blocks, the total computational complexity of the AO procedure over $N_{\mathrm{out}}$ outer iterations is
\begin{equation}    
\mathcal{O}\!\left(
N_{\mathrm{out}}K
\log\!\left(
\max\!\left(
\frac{\max_i M^{(i)}}{\delta_M},
\frac{\max_i P^{(i)}}{\delta_P},
\frac{\max_i \nu^{(i)}}{\delta_\nu}
\right)
\right)
\right).
\label{eq:complexity}
\end{equation}
Since $N_{\mathrm{out}}$ is typically small, the overall complexity scales linearly with the number of UEs and logarithmically with the desired numerical precision. This is significantly lower than that of an exhaustive grid search, whose complexity grows polynomially with the discretization of $P$, $M$, and $\boldsymbol{\omega}$.

\section{Simulation results}
\label{sec:res}
In this section, we evaluate the performance of the proposed {DEEP-DEAL} framework, considering both \textbf{Class B PA} and \textbf{Perfect PA}, under various scenarios and compare the results with certain benchmark algorithms, listed in Table \ref{tab: algos}. We compare our results with a reference scenario (REF-E), where the IBO $\Psi = 6$ dB is fixed, ensuring that the SDR is high enough to guarantee the mean Error Vector Magnitude (EVM) of $4.5\%$ required in $5$G New Radio for $256$-QAM constellation \cite{3gpp_38141}. Additionally, in one scenario we also consider a reference solution denoted as FixedSDR, commonly adopted in the literature \cite{6854179,9226127}, where the nonlinear distortion power is assumed to be proportional to the transmit power and the system operates in PA linear region, i.e., IBO equal 6 dB is assumed. This corresponds to a constant SDR. Under this assumption, the SNDR in (\ref{eq: SINR_ZF}) simplifies to
\begin{equation}
\gamma_k \approx \frac{(M-K)p_k\beta_k}{\sigma_k^2 + \beta_k \kappa \sum_k p_k},
\end{equation}
where $\lambda=1$ and $\kappa$ is a constant distortion factor. For fair comparison, $\kappa$ is computed using the soft-limiter model at $\Psi=6$ dB, resulting in $\kappa=0.0011$. Furthermore, the PA power consumption is approximated as $P_{\mathrm{PA}} \approx bP$, where $b$, derived from (\ref{eq:p_pa}) at $\Psi=6$ dB, equals $0.98$ and $2.24$ for a \textbf{Perfect PA} and \textbf{Class B PA}, respectively. Following \cite{9226127}, equal power allocation among UEs is assumed, while the IBO is optimized within $\Psi \in [3,9]$ dB. 
\begin{table}[!ht]
    \centering
    \begin{adjustbox}{width=1.0\columnwidth}
        \begin{tabular}{|>{\columncolor[HTML]{9B9B9B}}l|l|}
        \hline
        DEEP-DEAL &  Distortion Aware Energy Efficient Power Allocation  with \\ & Distortion Aware Energy Efficient Antenna Allocation   \\ \hline
        DEEP & Distortion Aware Energy Efficient Power Allocation with \\ & fixed Antenna Count \\ \hline
        REF-E & Reference Power (IBO equals 6 dB) with Equal per UE \\ & Power Distribution and with fixed Antenna Count  \\ \hline        
        FixedSDR & Reference solution assuming distortion power \\ 
        & proportional to transmit power (constant SDR model) \\ \hline
        \end{tabular}
    \end{adjustbox}
\caption{Tested algorithms abbreviation summary. }
\label{tab: algos}
\end{table}

The simulation parameters adopted in this work are shown in Table \ref{tab: sys_para}. It should be noted that in this section path loss in dB is denoted as $(\beta)_{\mathrm{dB}}$, which is related to the linear channel gain $\beta$ used in the previous sections as $\left( \beta \right)_{\mathrm{dB}}=-10\log_{10}(\beta)$.
\begin{table}[!ht]
\renewcommand{\arraystretch}{1.2}
    \centering
    \begin{adjustbox}{width=1.0\columnwidth}
        \begin{tabular}{|c|c|c|c|}
        \hline
        \rowcolor[HTML]{9B9B9B}
         $N_{\mathrm{U}}$ & $\Delta f$ & $\eta$ & $(\sigma^2)_{dBm}$  \\ \hline
        $1200$ & $15$ kHz & $\frac{2}{3}$ & $-174 \frac{dBm}{Hz} + 10  \cdot \log_{10}(N_{\mathrm{U}} \cdot \Delta f)$ \\ \hline
        \rowcolor[HTML]{9B9B9B}
        $P_{\mathrm{const}}$ \cite{10341276} & $P_{\mathrm{SPRF}}$ \cite{10341276} & $P_{\textrm{max}}$ & $(\beta_k)_{\mathrm{dB}}$ \cite{itu-m2135-1-2009}   \\ \hline
        $348$ W & $23$ W & $160$ W & $22.7 + 36.7 \cdot \log_{10}(d_k) + 26 \cdot \log_{10}(f_c)$  \\ \hline  
        \end{tabular}
    \end{adjustbox}
\caption{Simulation parameters,  where $d_k$ is the distance between UE $k$ and the base station in meters and $f_c =  3$ GHz \cite{itu-m2135-1-2009},  and $\delta_P = \delta_M = \delta_{\nu} = \delta_{EE} = 10 ^{-6}$. 
}
\label{tab: sys_para}
\end{table}

\subsection{Two-UE Homogeneous Path Loss}
\label{subsec:2_ue_homo}

We begin by examining the EE performance for the case of $K=2$ UEs experiencing identical large-scale fading conditions, which is possible when the UEs are at the same distance from the BS but on various azimuth angles. The DEEP and REF-E schemes employ a fixed antenna count of $M=32$, while DEEP--DEAL jointly optimizes both transmit power and antenna count. The considered path-loss range of $60$-$200$~dB spans a wide range of propagation conditions, from short-range links to extremely severe attenuation.

\begin{figure}[ht]
    \centering
    \includegraphics[width=0.80\columnwidth]{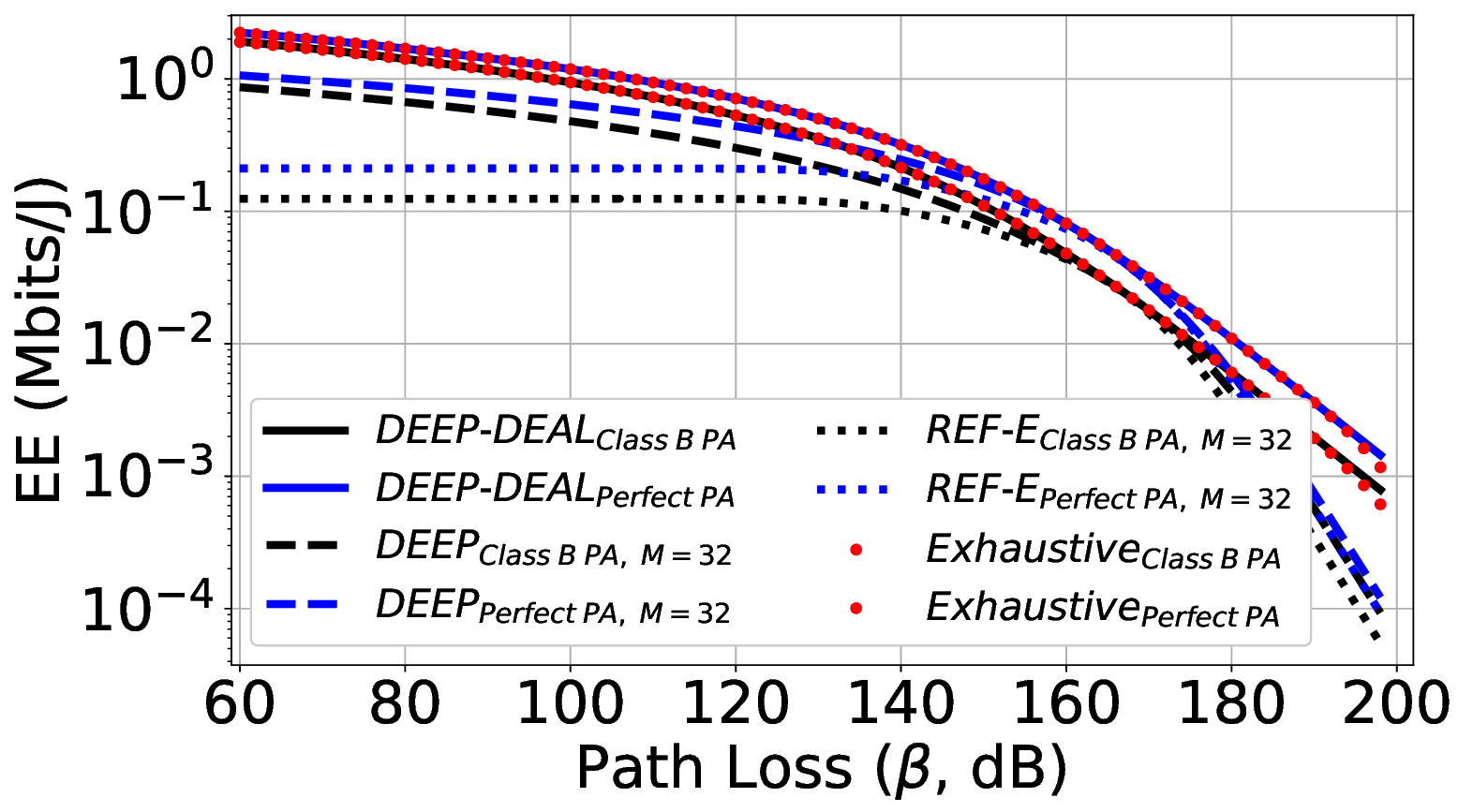}
    \caption{EE over varying path loss. }
    \label{fig:EE_opt}
\end{figure}

Fig.~\ref{fig:EE_opt} shows the resulting EE for all considered schemes under both PA architectures. As expected, EE decreases with increasing path loss for all algorithms due to worsening propagation conditions. The proposed {DEEP-DEAL} framework consistently achieves the highest EE across the entire range. In particular, substantial improvements over the fixed-IBO reference {REF-E} are observed at both short and long distances. For instance, at $\beta_{\mathrm{dB}}=80$ the EE improvement reaches about $50\%$ for the \textbf{Class B PA}, while for very short links the gain exceeds one order of magnitude since {DEEP-DEAL} activates only the minimum number of antennas and thus significantly reduces circuit power consumption.

A moderate-gap region appears around $\beta_{\mathrm{dB}}\approx160$--$170$~dB, where the EE differences between the considered schemes become smaller. In this region the fixed operating point used by {REF-E} happens to be closer to the EE-optimal PA operating point than in other path-loss conditions. 
However, it remains suboptimal since the configuration selected by {DEEP--DEAL} employs slightly different number of antennas and an IBO different than $6$~dB.   
For larger path loss values, the fixed-IBO assumption becomes increasingly inefficient, whereas the proposed algorithms continue adapting the operating point and maintain higher EE.

\begin{figure}[ht]
    \centering
    \includegraphics[width=0.80\columnwidth]{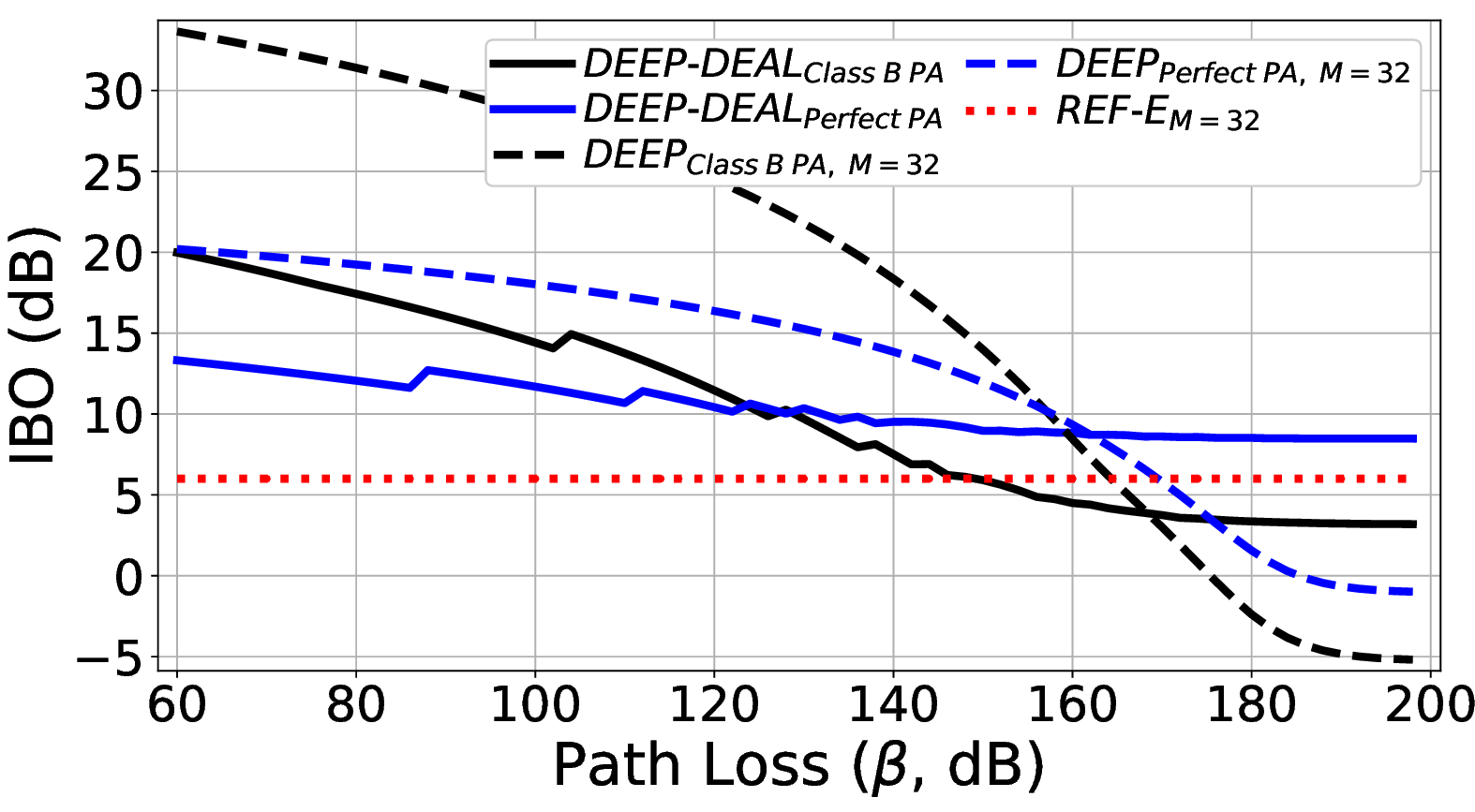}
    \caption{IBO over varying path loss.}
    \label{fig:IBO_opt}
\end{figure}
The advantage of {DEEP-DEAL} lies in its ability to jointly optimize transmit power and antenna count, which together determine the effective PA operating point through the IBO definition. In contrast, {DEEP} optimizes only transmit power while keeping $M$ fixed, and {REF-E} fixes the IBO a priori. Consequently, the additional antenna adaptation in {DEEP--DEAL} provides further EE improvements, particularly in short-link conditions where minimizing the number of active RF chains significantly reduces circuit power consumption.

To interpret the selected operating points, Fig.~\ref{fig:IBO_opt} shows the corresponding IBO values. For the {DEEP} algorithm, where $M$ is fixed, the optimal IBO decreases with increasing path loss as higher transmit power becomes necessary. Under the joint optimization of {DEEP-DEAL}, however, the IBO evolution becomes non-monotonic due to the coupling between transmit power and antenna configuration.

\begin{figure}[ht]
    \centering
    \includegraphics[width=0.80\columnwidth]{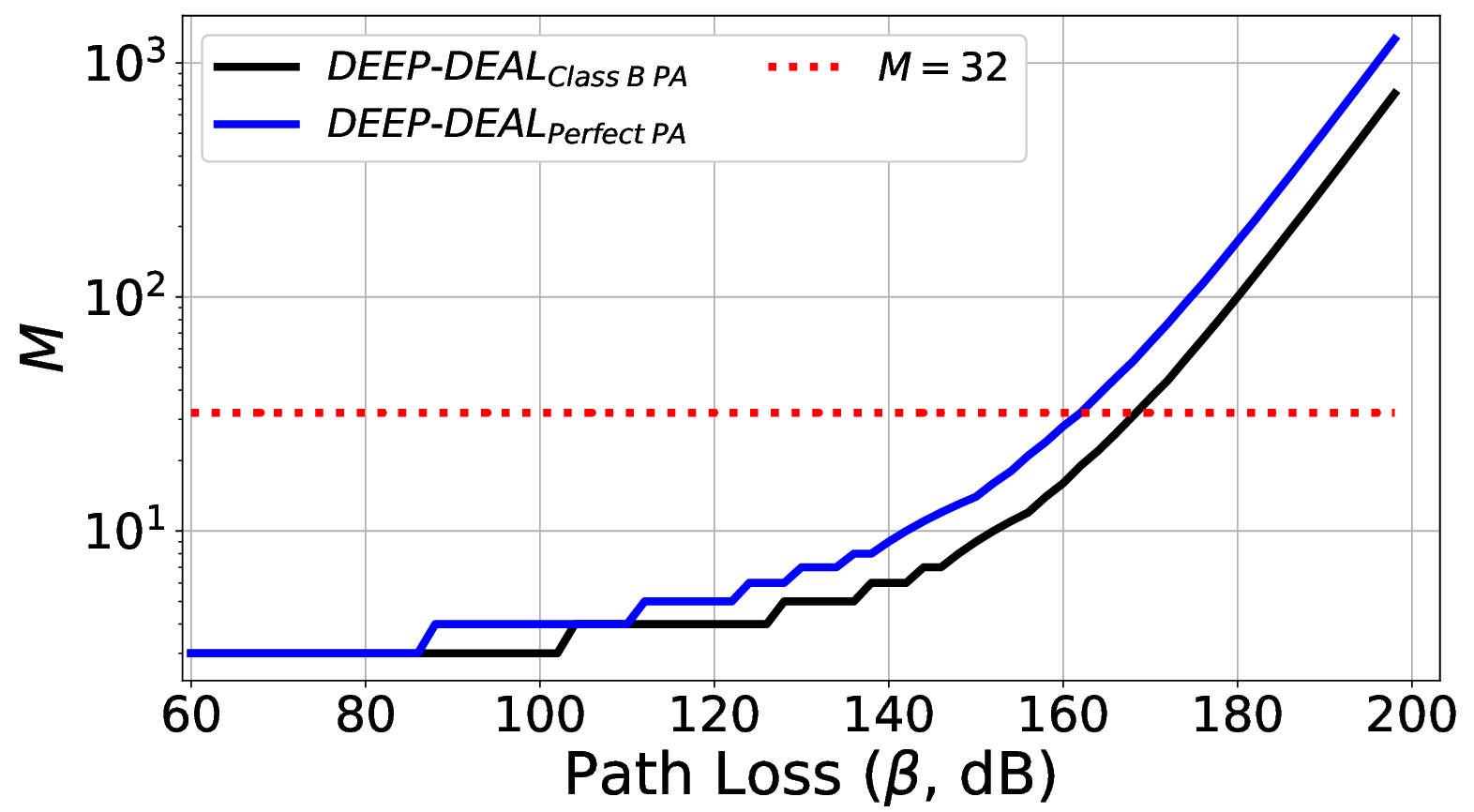}
    \caption{Optimal antenna count over varying path loss. }
    \label{fig:M}
\end{figure}

Fig.~\ref{fig:M} illustrates the corresponding optimal antenna count. For very short links, {DEEP-DEAL} selects the minimum feasible antenna count $M=3$, required for ZF precoding with $K=2$ UEs. As path loss increases, activating additional antennas becomes beneficial due to the ZF array gain scaling with $(M-K)$, leading to a progressive increase in the optimal antenna count. Systems with \textbf{Perfect PAs} typically activate more antennas than those with \textbf{Class B PAs}. This behavior arises because Perfect PAs do not incur an efficiency penalty when increasing transmit power, making antenna array gain more attractive for EE, whereas Class B PAs can improve EE by operating at lower IBO with fewer RF chains. The discrete antenna changes explain the non-monotonic behavior observed in the IBO curves in Fig.~\ref{fig:IBO_opt}, which follows directly from the IBO definition: increasing the antenna count raises $\Psi$, whereas higher transmit power reduces it.

Finally, Fig.~\ref{fig:EE_opt} also reports the results obtained via exhaustive search over the joint space $(M,P)$ with $M\in\{K+1,\ldots,500\}$ and $P\in\{10,11,\ldots,15000\}$~W. Since the scenario is homogeneous, the optimal power fractions satisfy $\omega_k=1/K$ and do not require exhaustive evaluation. The results confirm that {DEEP-DEAL} consistently achieves the global optimum while requiring orders of magnitude fewer evaluations as discussed in Sec. \ref{subsec:complexity}.

\subsection{Two-UE Heterogeneous Path Loss}
\label{subsec:2_ue_hetero}
\begin{figure}[ht]
    \centering
    \includegraphics[width=0.8\columnwidth]{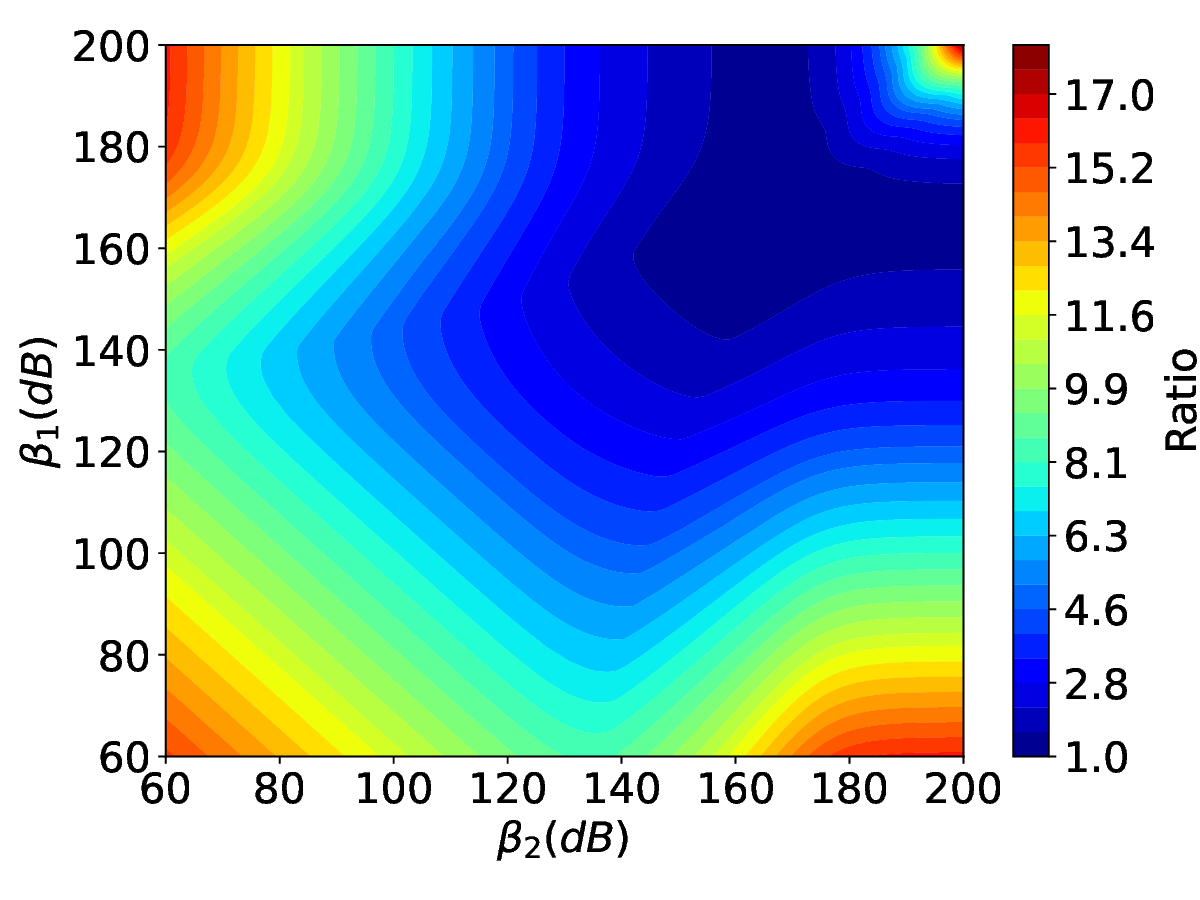}
    \caption{\textbf{Class B PA} - Ratio of EE obtained via {DEEP-DEAL} and REF-E ($M = 32$, $\Psi=6~dB$) for $2$ UEs over varying path loss scenario.}
    \label{fig:ratio_comp_opt_REFE_B}
\end{figure}

\begin{figure}[ht]
    \centering
    \includegraphics[width=0.8\columnwidth]{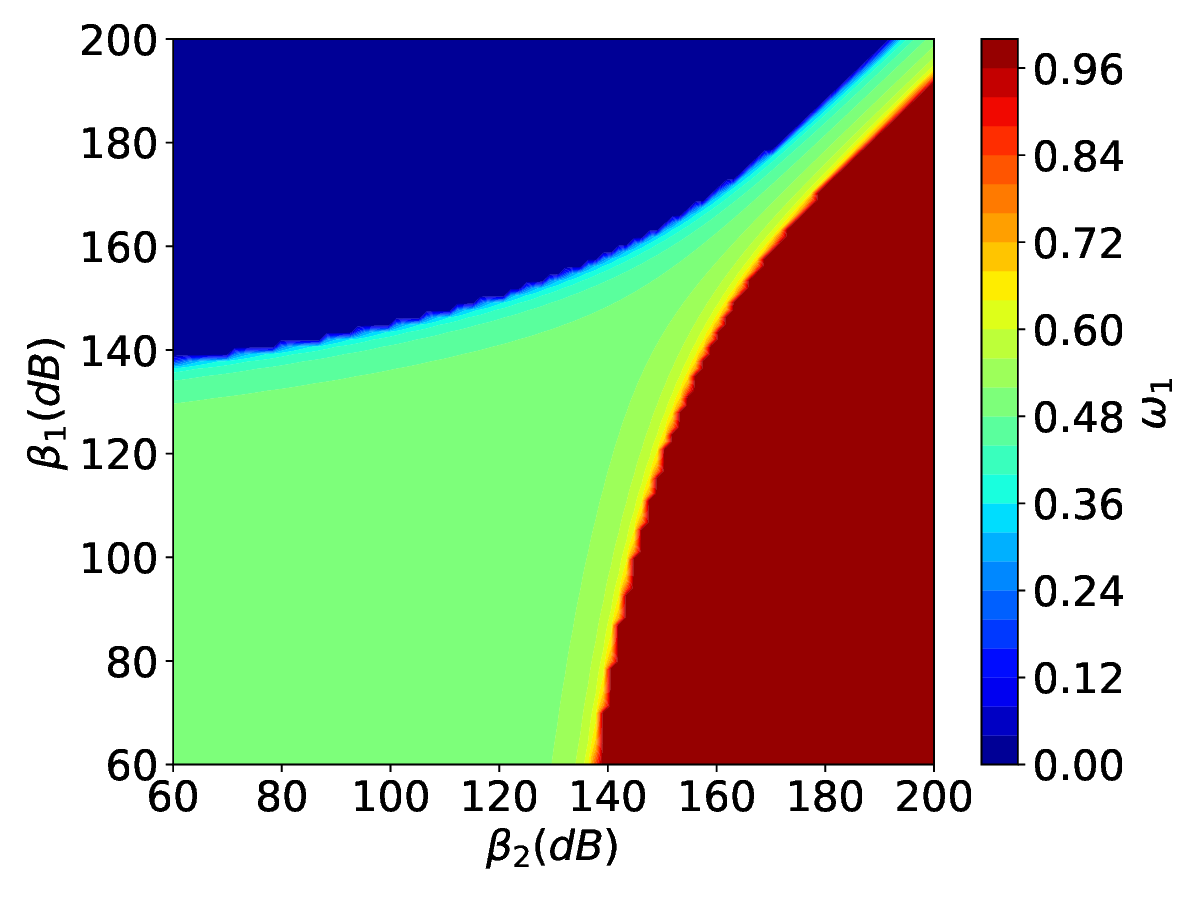}
    \caption{\textbf{Class B PA} - Fraction of power allocated by {DEEP-DEAL} to UE $1$ ($\omega_1$) for a $2$ UE scenario.}
    \label{fig:omega_1_opt_B}
\end{figure}

\begin{figure}[ht]
    \centering
    \includegraphics[width=0.8\columnwidth]{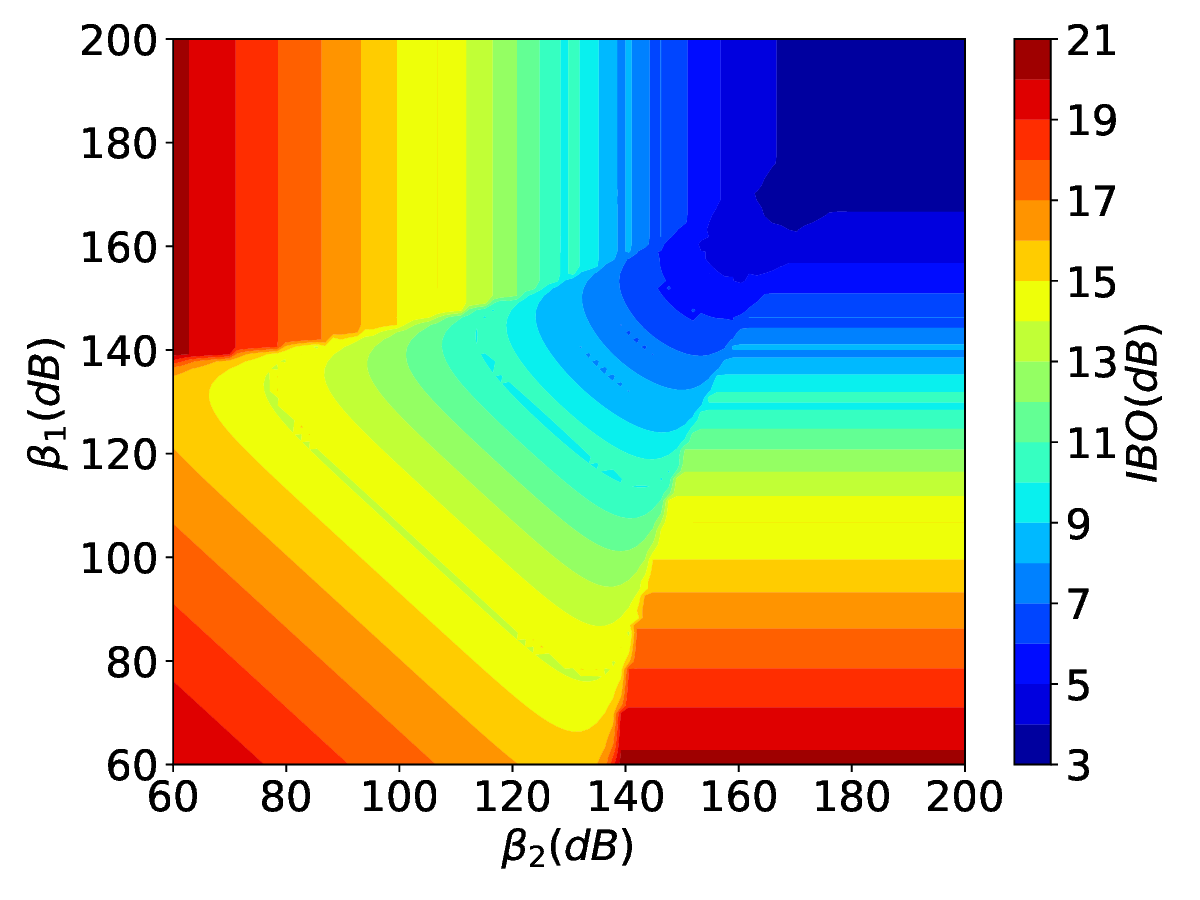}
    \caption{\textbf{Class B PA} - Optimum IBO by {DEEP-DEAL} for a $2$ UE scenario.}
    \label{fig:opt_IBO_B}
\end{figure}

\begin{figure}[ht]
    \centering
    \includegraphics[width=0.8\columnwidth]{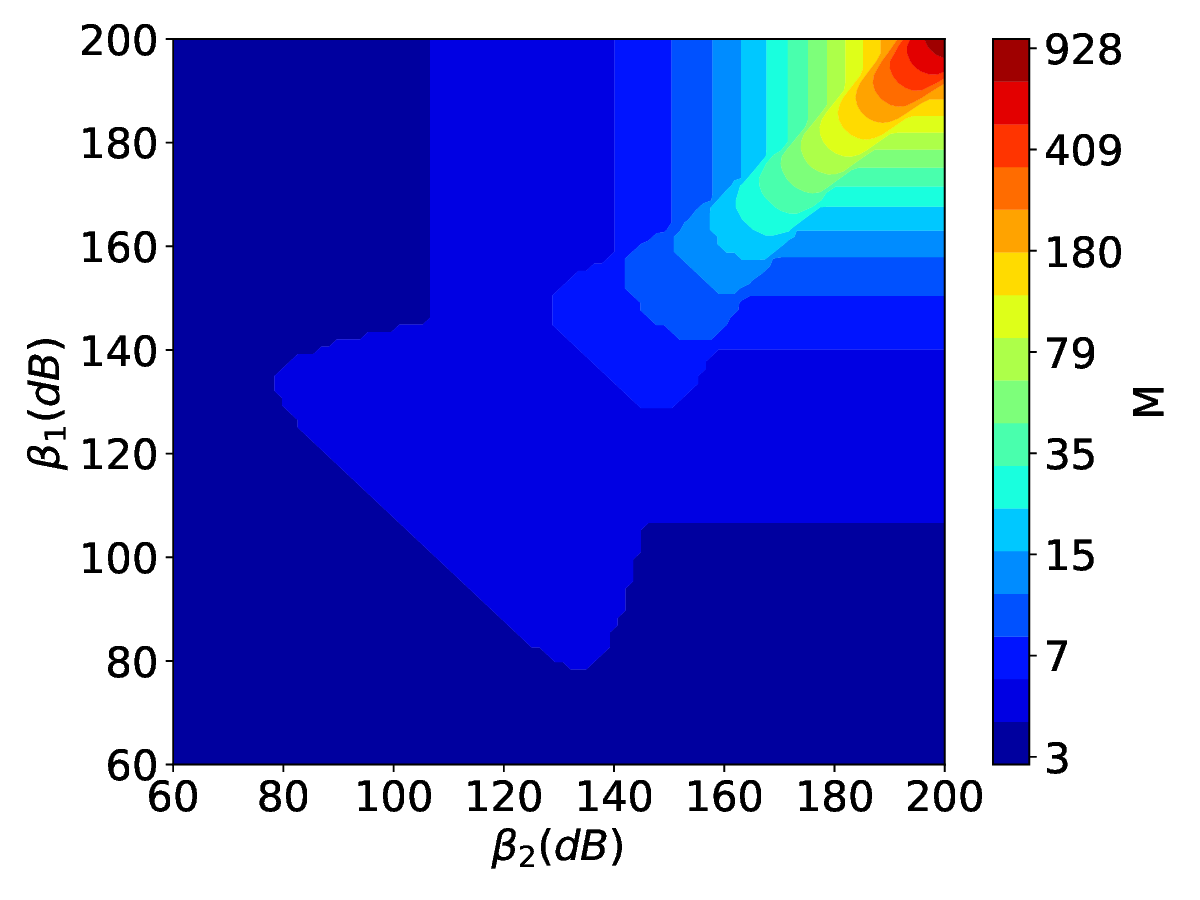}
    \caption{\textbf{Class B PA} - Optimum integer M by {DEEP-DEAL} for a $2$ UE scenario.}
    \label{fig:opt_M_B}
\end{figure}

While the previous subsection considered homogeneous propagation conditions, we now investigate heterogeneous channel scenarios where UEs experience different path losses.
As depicted in Figs.~\ref{fig:ratio_comp_opt_REFE_B}-\ref{fig:opt_M_B}, we next examine the $2$-UE heterogeneous scenario, where each UE experiences a different path loss $\beta$. The path-loss values are varied on a $2$-D grid from $60$ to $200$~dB. The diagonal corresponds to the homogeneous case discussed in Sec.~\ref{subsec:2_ue_homo}, whereas the off-diagonal regions represent heterogeneous propagation conditions.

Fig.~\ref{fig:ratio_comp_opt_REFE_B} shows the ratio between the EE achieved by {DEEP-DEAL} and the REF-E allocation. The proposed framework outperforms REF-E in nearly all scenarios. Only a small region along the diagonal at relatively high path-loss values exhibits little or no improvement, corresponding to the operating region where the EE curves in Fig.~\ref{fig:EE_opt} approach each other in the homogeneous case. Outside this region, substantial gains appear. In particular, when at least one UE is close to the BS, the EE improvement reaches up to one order of magnitude compared with the REF-E allocation. The symmetry of the map indicates that the performance mainly depends on the relative channel strengths rather than on the UE indices.

The mechanisms behind these gains can be understood by jointly examining the optimal power allocation, antenna activation, and the resulting PA operating point. As shown in Fig.~\ref{fig:omega_1_opt_B}, a funnel-shaped region of equal power allocation ($\omega_1=\omega_2=0.5$) appears when both UEs experience comparable propagation conditions. This quasi-homogeneous regime extends well beyond the diagonal, for example reaching $(\beta_1,\beta_2)=(100,70)$~dB. In this region equal power allocation is EE-optimal, while the optimal IBO and antenna count exhibit structured stripe-like patterns perpendicular to the diagonal, as visible in Figs.~\ref{fig:opt_IBO_B} and~\ref{fig:opt_M_B}. This indicates that the EE-optimal operating point is primarily governed by the aggregate path loss rather than the exact imbalance between the UEs. Consistent with the homogeneous case, the optimal IBO decreases with increasing path loss, whereas the number of active antennas increases. The sharp transitions observed in the IBO map arise from the integer-valued antenna constraint imposed in the final step of Algorithm~\ref{alg:ao}. The emergence of the equal-power funnel further highlights that moderate channel asymmetry does not necessarily require unequal power allocation, since similar propagation conditions lead to nearly identical EE-optimal operating points for both UEs.

Outside the equal-power funnel the EE-optimal solution allocates most of the transmit power to the UE with the stronger channel. Consequently, the antenna configuration and the PA operating point are largely determined by this UE. This behavior reflects the EE-optimal strategy of prioritizing the UE with the more favorable channel conditions. As the path loss of the stronger UE increases, the optimal IBO generally decreases while the number of active antennas increases to compensate for the propagation losses. Minor non-monotonic variations in the IBO remain visible due to the discrete antenna adaptations within the joint $P$--$M$ optimization.

\begin{figure}[ht]
    \centering
    \includegraphics[width=0.8\columnwidth]{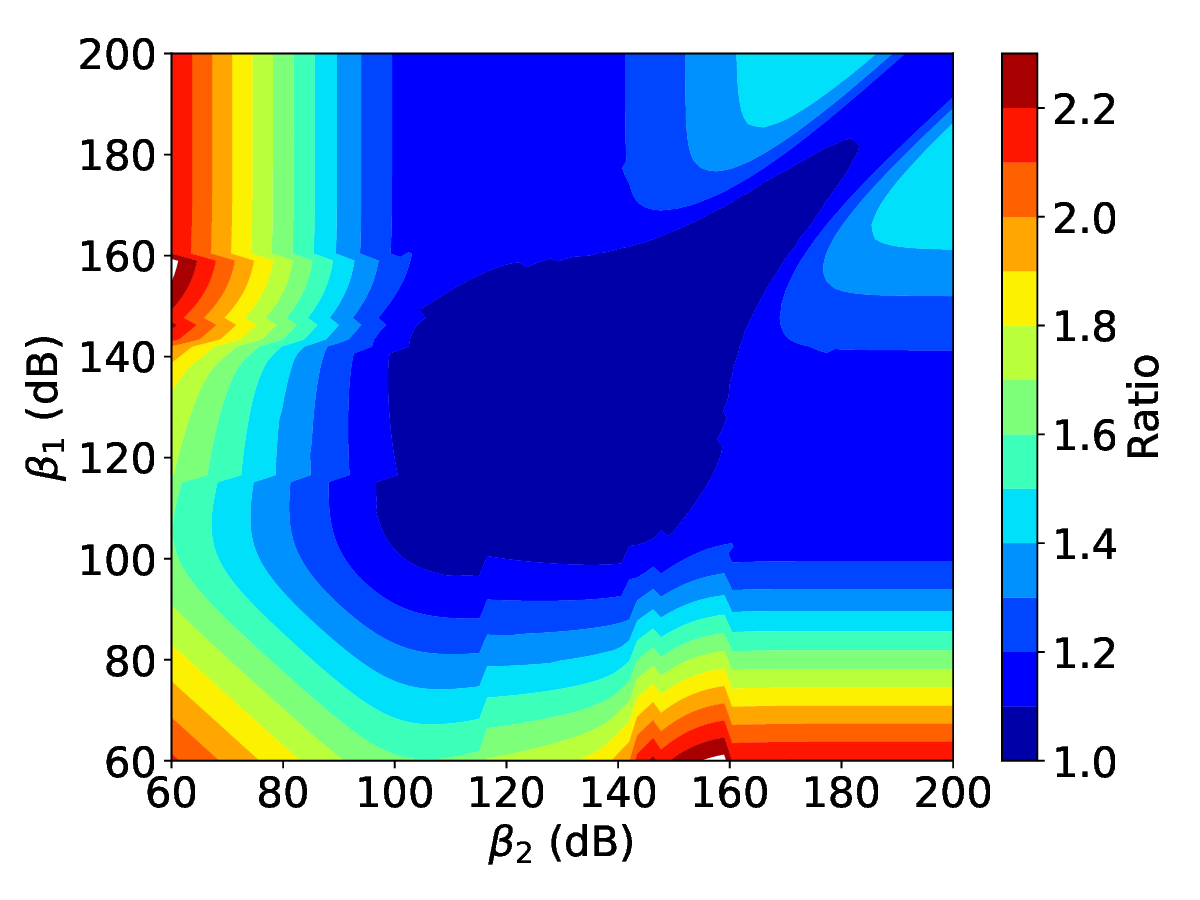}
    \caption{\textbf{Class B PA} - Ratio of EE obtained via {DEEP-DEAL} and FixedSDR for $2$ UEs over varying path loss with $M = 32$ and IBO optimized within $\Psi \in [3,9]$ dB.}
    \label{fig:ratio_comp_opt_kappa_B}
\end{figure}
For completeness, Fig.~\ref{fig:ratio_comp_opt_kappa_B} compares the proposed framework with the FixedSDR approximation, where nonlinear distortion is modeled as proportional to the transmit power. The ratio surface shows that both approaches perform similarly in a broad central region of the $(\beta_1,\beta_2)$ plane, where the optimal operating point remains close to the reference IBO of 6 dB used to derive the FixedSDR model. However, as the channel conditions become more asymmetric or the optimal IBO deviates from this reference, the performance gap increases. In moderately asymmetric scenarios the proposed distortion-aware solution provides EE gains of roughly $20$–$40\%$, while in extreme regimes the improvement can exceed $2\times$. The gains become particularly pronounced when the UEs experience highly asymmetric path losses, where most transmit power is allocated to the stronger UE and the optimal PA operating point deviates significantly from the reference IBO used in the FixedSDR approximation. These results highlight that fixed-SDR approximation, commonly used in the literature, remains locally accurate but may significantly underestimate the benefits of jointly optimizing the PA operating point and antenna configuration under extreme channel conditions.

\subsection{Multi-UE Heterogeneous Path Loss}
\label{subsec:multi_ue}
\begin{figure}[ht]
    \centering

    \begin{subfigure}{0.80\columnwidth}
        \centering
        \includegraphics[width=\linewidth]{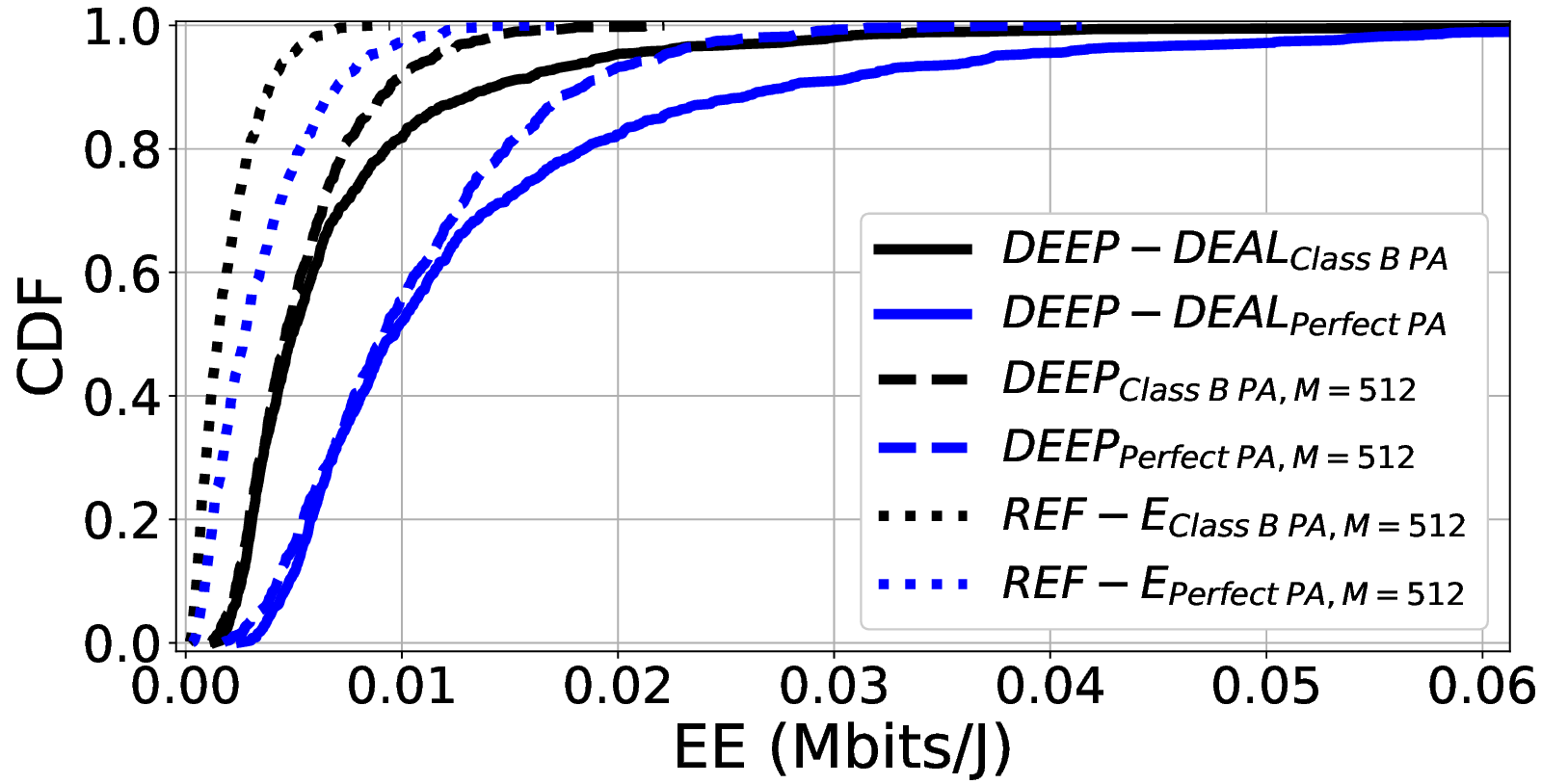}
        \caption{EE CDF.}
        \label{fig:multi_ue_EE}
    \end{subfigure}


    \begin{subfigure}{0.80\columnwidth}
        \centering
        \includegraphics[width=\linewidth]{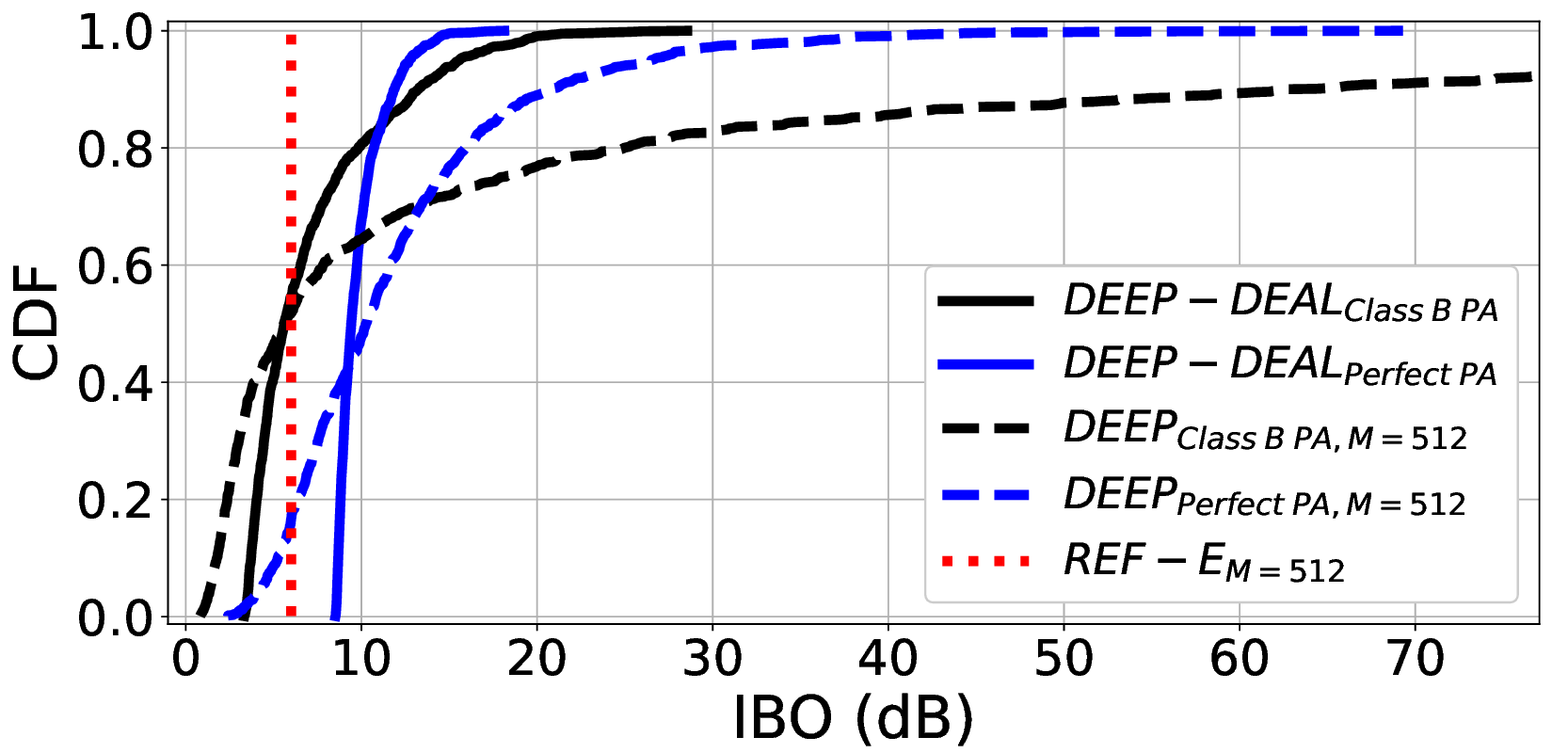}
        \caption{IBO CDF.}
        \label{fig:multi_ue_IBO}
    \end{subfigure}


    \begin{subfigure}{0.80\columnwidth}
        \centering
        \includegraphics[width=\linewidth]{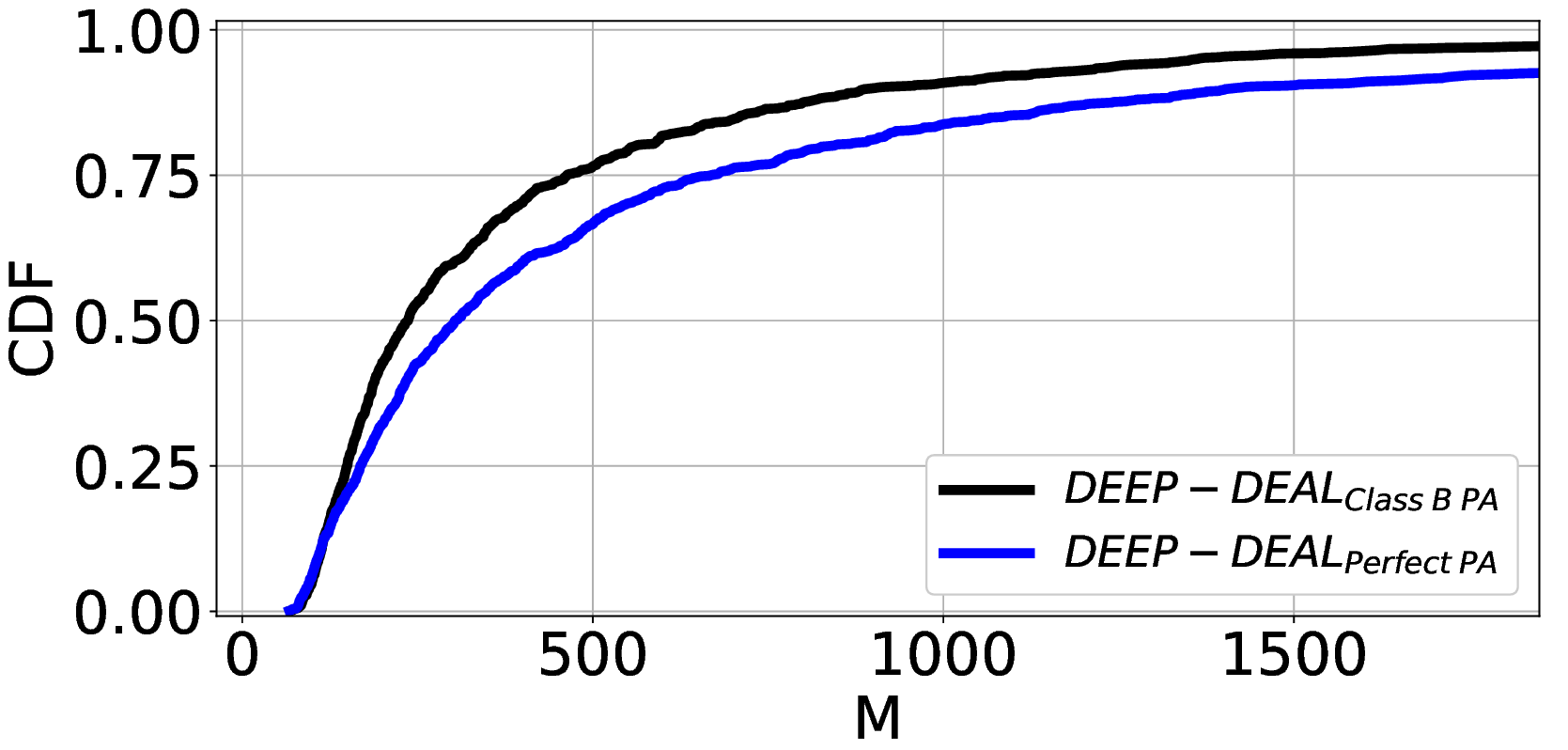}
        \caption{Number of antennas $M$ CDF.}
        \label{fig:multi_ue_M}
    \end{subfigure}

    \caption{CDF performance metrics in a heterogeneous scenario with 60 UEs uniformly distributed in a 10 km radius cell. }
    \label{fig:multi_ue_combined}
\end{figure}
Finally, to evaluate the behavior of the proposed DEEP-DEAL framework under realistic network conditions, we consider a large multi-UE scenario. A single circular cell of radius $10$~km is assumed, where $K=60$ UEs are uniformly distributed over the cell area. For each experiment, $1000$ random UE location realizations are generated according to the path-loss model in Table~\ref{tab: sys_para}.

Figs.~\ref{fig:multi_ue_EE} and \ref{fig:multi_ue_IBO} present the cumulative distribution functions (CDFs) of the achieved EE and the corresponding effective IBO for the considered algorithms under both PA architectures. The proposed {DEEP-DEAL} framework consistently outperforms all reference schemes across the entire CDF range, with the largest improvements observed in the high-EE region. At the median, the EE improvement relative to the REF-E allocation reaches approximately a threefold gain, while improvements relative to the DEEP baseline remain around 40--50\%. These improvements are achieved through the joint adaptation of transmit power, antenna count, and per-UE power allocation.

The corresponding IBO distributions reveal substantial variability across UE realizations and PA architectures. When antenna adaptation is disabled, as in the {DEEP} scheme, the algorithm compensates for unfavorable propagation conditions by significantly increasing the PA operating point, leading to very large IBO values. For visualization clarity, Fig.~\ref{fig:multi_ue_IBO} shows a zoomed view of the CDF to highlight the differences between the schemes, since the curves converge toward one for large IBO values. In contrast, {DEEP-DEAL} exploits antenna adaptation to maintain a more balanced PA operating regime while still achieving higher EE.

Fig.~\ref{fig:multi_ue_M} illustrates the distribution of the EE-optimal number of active antennas obtained by {DEEP-DEAL}. Although a minimum of $M=61$ antennas is required to serve $60$ UEs with ZF precoding, the EE-optimal antenna count is typically much larger due to the increased path-loss variability in the large-cell scenario. In particular, the median antenna count reaches several hundred antennas, indicating that large antenna arrays are beneficial for maintaining high EE in dense multi-user deployments when realistic PA characteristics are considered.

\subsection{Convergence of Algorithm \ref{alg:ao}}
\label{subsec:convg_plot}
\begin{figure}[ht]
    \centering
    \includegraphics[width=0.80\columnwidth]{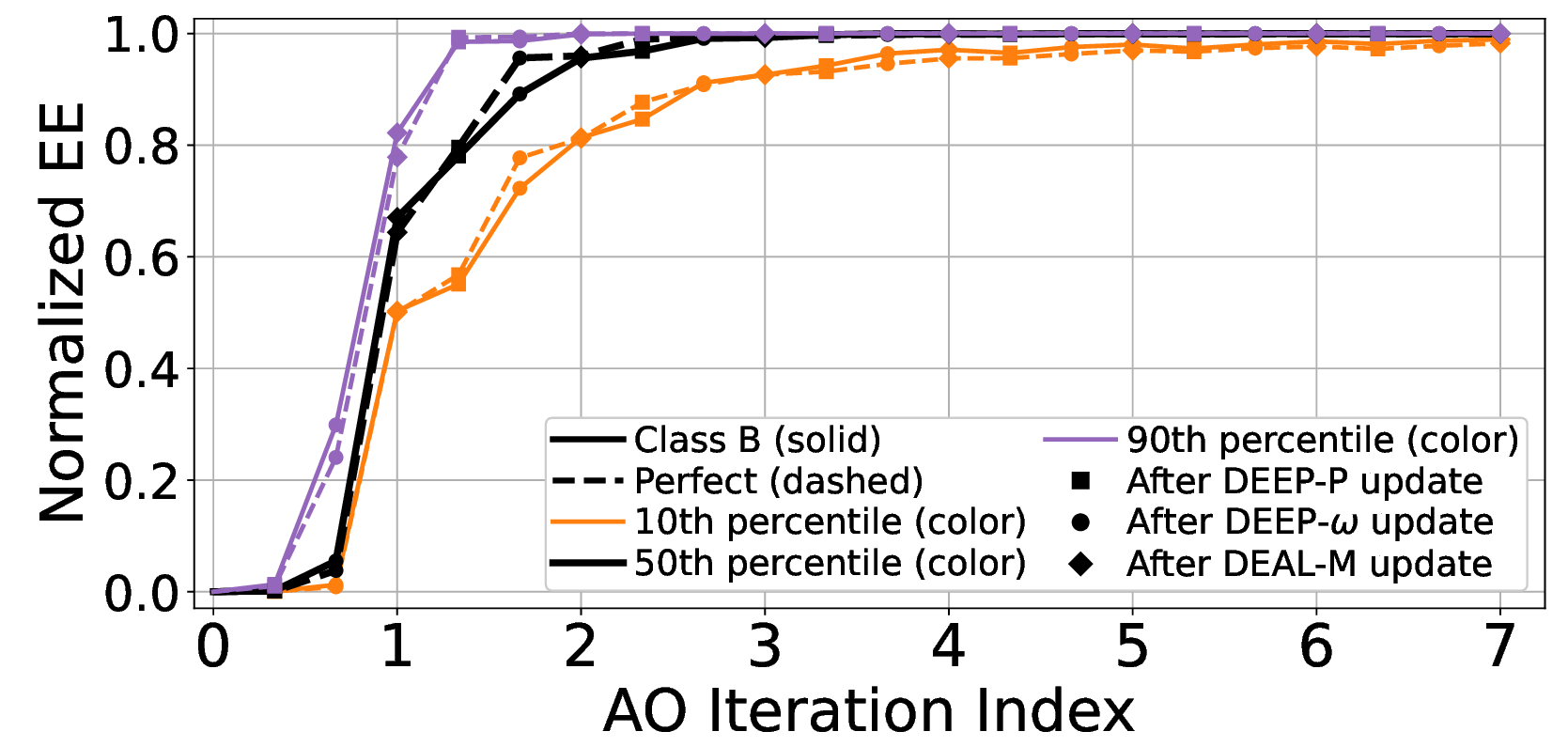}
    \caption{Empirical convergence of the proposed AO Algorithm \ref{alg:ao} shown via the trajectory through $P \to \boldsymbol{\omega} \to M$ updates.     }
    \label{fig:conv_plt}
\end{figure}

Fig.~\ref{fig:conv_plt} illustrates the convergence behavior of the proposed Algorithm \ref{alg:ao} for both Class~B and Perfect PA models. The results are obtained over 1000 independent channel realizations in a heterogeneous 60-UE scenario, where the EE of each realization is normalized by its final value. The figure confirms the theoretical monotonicity established in Section~IV: the EE is monotonically non-decreasing and stabilizes rapidly.

The normalized EE is plotted versus the AO iteration index, where fractional indices correspond to intermediate updates within a single AO cycle. Importantly, monotonic ascent is observed not only across outer AO iterations but also across the internal block updates. Solid and dashed curves correspond to the Class~B and Perfect PA models, respectively, while different colors indicate the 10th, 50th, and 90th percentiles across channel realizations, capturing convergence variability.

Quantitatively, defining convergence as achieving at least $99.9\%$ of the final EE, the median number of outer AO iterations is 5 for the Class~B PA and 4 for the Perfect PA. More than $90\%$ of realizations converge within 7 iterations, while even the slowest cases require no more than 13 and 18 iterations, respectively. Similar convergence rates are observed for the individual block updates, indicating balanced progress across optimization variables. Overall, the results demonstrate fast and robust convergence of the proposed AO framework across PA models and channel conditions.

\section{Conclusion and Future Work}
\label{sec:conc}

In this work, we first modeled the impact of soft-limiter PA characteristics on the EE of a MU-M-MIMO OFDM DL system employing ZF precoding under an i.i.d. Rayleigh channel. We defined the resource allocation problem and proposed the {DEEP-DEAL} framework, which jointly optimizes distortion-aware power control (DEEP) and antenna allocation (DEAL) in an alternating optimization manner. Unlike conventional formulations, the optimization problem does not require explicit transmit power constraints, as the nonlinear PA inherently limits the feasible operating region through distortion. Our simulation results show that {DEEP-DEAL} consistently outperforms fixed allocation and power-only optimizing (DEEP) baselines across homogeneous and heterogeneous UE scenarios and across PA classes. Notably, the optimum frequently operates at low IBO for high path loss cases, while jointly adapting the number of active antennas. This indicates that operating the PA near saturation is EE-optimal when paired with antenna scaling. Overall, the findings underscore that balancing distortion-aware power control with antenna scaling is essential for maximizing EE in M-MIMO systems.

Several promising research directions emerge from this study. First, while {DEEP-DEAL} was evaluated for ZF precoding, extending the framework to other schemes such as MRT or hybrid precoding would provide further insights into its applicability under different interference and complexity trade-offs. Second, incorporating multi-cell scenarios with inter-cell interference and coordinated antenna activation could more closely reflect practical deployment conditions. Third, the present work assumes perfect CSI that can be extended to consider the channel estimation errors. Additionally, the optimization framework can be enriched by integrating traffic-aware or QoS-driven objectives, allowing energy-efficient adaptation to varying service demands. Finally, the hardware model can be extended, e.g., considering the envelope tracking PA, or distortions coming from different front-end elements like the DAC.

\appendices

\section{Important Limits and Expansions}
\label{sec:approxis}

In the following appendices we employ the limits \cite{abramowitz1964}
\begin{align}
&\text{L1}: \lim_{x\to \infty} x^ae^{-x} = 0 \quad \forall a \in \mathbb{R}, \quad
\text{L2}: \lim_{x \to \infty} \mathrm{erf}(x) = 1, \nonumber\\
&\text{L3}: \lim_{x\to \infty} x^a\mathrm{erfc}(x) = 0 \quad \forall a \in \mathbb{R}
\end{align}
and the asymptotic expansions as $x \to 0$
\begin{align}
&\text{A1}: \log_2(1+x) \sim \frac{x}{\ln(2)}, \quad
\text{A2}: e^{-x} \sim 1 - x + \frac{x^2}{2}, \nonumber\\
&\text{A3}: \mathrm{erfc}(x) \sim
1 - \frac{2x}{\sqrt{\pi}} + \frac{2x^3}{3\sqrt{\pi}}, \quad
\text{A4}: \mathrm{erf}(x) \sim \frac{2x}{\sqrt{\pi}} - \frac{2x^3}{3\sqrt{\pi}}.
\end{align}
Throughout the appendix we use the asymptotic equivalence notation $\sim$ \cite{tse2005fundamentals}.

\section{Proof of Lemma \ref{sec:proof_lemma_1}}
\label{sec:proof_lemma_1}

The derivative of the EE with respect to $P$ is
\begin{align}
\frac{\partial \mathrm{{EE}}}{\partial P} =
\frac{\sum_k\frac{\partial R_k}{\partial P} P_{\mathrm{tot}}
- \sum_k R_k \frac{\partial P_{\mathrm{tot}}}{\partial P}}
{P_{\mathrm{tot}}^2}.
\label{eq:ee_d}
\end{align}
Since $P_{\mathrm{tot}}>0$, the stationary point is determined by the numerator of \eqref{eq:ee_d}. This has the same root as
\begin{equation}
f_P(P) =
\frac{1}{\sum_k R_k}\sum_k \frac{\partial R_k}{\partial P}
-
\frac{1}{P_{\mathrm{tot}}}\frac{\partial P_{\mathrm{tot}}}{\partial P}.
\label{eq:der_EE}
\end{equation}

Let $B=N_U\Delta f$. With $\gamma_k=\frac{(M-K)\lambda\omega_k P\beta_k}{\sigma_k^2+\beta_k D}$, the derivative of the rate equals
\begin{align}
\frac{\partial R_k}{\partial P}=&
\frac{B(M-K)}{\ln(2)(1+\gamma_k)}
\frac{\omega_k\beta_k}{(\sigma_k^2+\beta_k D)^2}
\nonumber\\
&\cdot
\left(
(\partial_P\lambda\,P+\lambda)(\sigma_k^2+\beta_k D)
-\lambda P\beta_k\partial_P D
\right),
\label{eq:der_Rk}
\end{align}
where
\begin{align}
\frac{\partial \lambda}{\partial P}=
-\frac{\Psi\sqrt{\lambda}}{P}
\left(
e^{-\Psi}+\frac{1}{2}\sqrt{\frac{\pi}{\Psi}}\mathrm{erfc}(\sqrt{\Psi})
\right),
\label{eq:der_lambda}
\end{align}
\begin{align}
\frac{\partial D}{\partial P}=
\eta\left(1-e^{-\Psi}-\lambda-P\frac{\partial\lambda}{\partial P}-\Psi e^{-\Psi}\right),
\label{eq:der_D}
\end{align}
with $\Psi=MP_{\max}/P$.
Furthermore,
\begin{equation}
\frac{\partial P_{\mathrm{tot}}}{\partial P}
=\frac{\partial P_{\mathrm{PA}}}{\partial P}
\end{equation}
since other components of $P_{\mathrm{tot}}$ are independent of $P$, where
\begin{equation}
\frac{\partial P_{\mathrm{PA}}}{\partial P} =
\begin{cases}
\dfrac{\sqrt{\Psi}}{\sqrt{\pi}}\mathrm{erf}(\sqrt{\Psi})-\dfrac{2}{\pi}\Psi e^{-\Psi}, & \textbf{Class B PA}\\
1-e^{-\Psi}-\Psi e^{-\Psi}, & \textbf{Perfect PA}.
\end{cases}
\label{eq:der_p_tot}
\end{equation}

To analyze the existence of a root of $f_P(P)$, we examine the limits as $P\to0$ and $P\to\infty$.

\textbf{Case 1 ($P\to0$):}
In this regime $\Psi\to\infty$. Using L1 and L3 yields $\lambda\to1$ and $D\to0$. With $P=MP_{\max}/\Psi$,
\begin{equation}
\gamma_k \sim \frac{M(M-K)\omega_kP_{\max}\beta_k}{\sigma_k^2\Psi}.
\label{eq_simp_gamma}
\end{equation}

Since $\gamma_k\to0$, using A1 gives
\begin{equation}
R_k \sim
\frac{BM(M-K)P_{\max}\omega_k\beta_k}{\ln(2)\sigma_k^2\Psi}
\end{equation}
and
\begin{equation}
\sum_k R_k \sim
\frac{BM(M-K)P_{\max}}{\ln(2)\Psi}
\sum_k\frac{\omega_k\beta_k}{\sigma_k^2}.
\label{eq:sum_r_k}
\end{equation}

Moreover $\partial_P\lambda\to0$ and $\partial_P D\to0$, giving
\begin{align}
\sum_k\frac{\partial R_k}{\partial P}
\sim
\frac{B(M-K)}{\ln(2)}
\sum_k\frac{\omega_k\beta_k}{\sigma_k^2}.
\end{align}

Using the asymptotic expansion of $\mathrm{erf}(x)$ for $x\to\infty$,
\begin{equation}
\mathrm{erf}(x)\sim
1-\frac{e^{-x^2}}{x\sqrt{\pi}}\left(1-\frac{1}{2x^2}\right),
\label{eq:asym_erf}
\end{equation}
(\ref{eq:der_p_tot}) simplifies to
\begin{align}
\frac{\partial P_{\mathrm{PA}}}{\partial P}\sim
\begin{cases}
\dfrac{\sqrt{\Psi}}{\sqrt{\pi}}, & \textbf{Class B PA}\\
1, & \textbf{Perfect PA}.
\end{cases}
\label{eq:der_pa}
\end{align}

Similarly,
\begin{align}
P_{\mathrm{tot}}\sim
\begin{cases}
P_{\mathrm{const}}+MP_{\mathrm{SPRF}}, & \textbf{Class B PA}\\
P+P_{\mathrm{const}}+MP_{\mathrm{SPRF}}, & \textbf{Perfect PA}.
\end{cases}
\label{eq_P_total_as_P_0}
\end{align}

Substituting (\ref{eq:sum_r_k})–(\ref{eq_P_total_as_P_0}) into (\ref{eq:der_EE}) yields
\begin{align}
f_P(P)\sim
\begin{cases}
\frac{1}{\sqrt{P}}\left(\frac{1}{\sqrt{P}}-\frac{\sqrt{MP_{\max}}}{\sqrt{\pi}P_{\mathrm{tot}}}\right), & \textbf{Class B PA}\\
\frac{1}{P}-\frac{1}{P+P_{\mathrm{const}}+MP_{\mathrm{SPRF}}}, & \textbf{Perfect PA}.
\end{cases}
\end{align}

Thus $f_P(P)\to+\infty$ as $P\to0$.

\textbf{Case 2 ($P\to\infty$):}
Here $\Psi\to0$. Using A2–A3 gives
\begin{equation}
\lambda \sim \frac{\pi}{4}\Psi,
\label{eq:lambda_approx}
\end{equation}
and
\begin{equation}
D \sim \eta\left(1-\frac{\pi}{4}\right)MP_{\max}.
\label{eq:D_approx}
\end{equation}

Defining $C_k=\sigma_k^2+\eta\beta_k(1-\frac{\pi}{4})MP_{\max}$,
\begin{align}
\gamma_k\sim
\frac{\pi M(M-K)\omega_k\beta_kP_{\max}}{4C_k}.
\label{eq:gamma_approx}
\end{align}
Hence $\sum_k R_k$ converges to a constant. Using A2–A3 in (\ref{eq:der_lambda}) and (\ref{eq:der_D}) yields
\begin{equation}
\frac{\partial\lambda}{\partial P}\sim-\frac{\pi\Psi}{4P}, \quad
\frac{\partial D}{\partial P}\sim\frac{\eta\Psi^2}{2}.
\label{eq:der_lambda_D}
\end{equation}

Substituting these expressions into (\ref{eq:der_Rk}) gives
\begin{align}
\sum_k\frac{\partial R_k}{\partial P}
\sim
-\sum_k
\frac{BM(M-K)}{8\ln(2)(1+\gamma_k)}
\frac{\eta\omega_k\beta_k^2\pi P_{\max}\Psi^2}{(\sigma_k^2+\beta_kD)^2}.
\end{align}

Using A2 and A4 in (\ref{eq:der_p_tot}) shows that
\begin{equation}
\frac{1}{P_{\mathrm{tot}}}\frac{\partial P_{\mathrm{tot}}}{\partial P}\to0.
\end{equation}

Thus
\begin{align}
    f_P(P) &\sim - \frac{1}{\sum_k\log_2(1+\gamma_k)} \cdot \nonumber \\
    &\sum_k
    \frac{BM(M-K)}{8\ln(2) \left( 
    1+\gamma_k
    \right)
    }
    \frac{\eta\omega_k\beta_k^2\pi P_{\mathrm{max}}  \Psi^2}{\left(\sigma_k^2+\beta_k D\right)^2}
    \label{eq:f_p_approx}
\end{align}
Since $\Psi\to0$, $f_P(P)\to0^-$. Therefore, $f_P(P)$ is positive as $P\to0$ and negative as $P\to\infty$. By the Intermediate Value Theorem, there exists at least one root in $(0,\infty)$ for both \textbf{Class B} and \textbf{Perfect PA}.

\bibliographystyle{IEEEtran}
\bibliography{IEEEabrv,bibtex/bib/IEEEexample}

\end{document}